\pdfoutput=1
\RequirePackage{ifpdf}
\ifpdf 
\documentclass[pdftex]{sigma}
\else
\documentclass{sigma}
\fi

\numberwithin{equation}{section}

\newcommand{\CA}{{\cal A}}

\newcommand{\CC}{{\cal C}}
\newcommand{\CE}{{\cal E}}
\newcommand{\CF}{{\cal F}}

\newcommand{\CM}{{\cal M}}
\newcommand{\CN}{{\cal N}}
\newcommand{\CO}{{\cal O}}

\newcommand{\CR}{{\cal R}}
\newcommand{\CS}{{\cal S}}

\newcommand{\CV}{{\cal V}}
\newcommand{\CW}{{\cal W}}

\newcommand{\CZ}{{\cal Z}}
\def\IZ{{\mathbb Z}}
\def\IR{{\mathbb R}}
\def\IC{{\mathbb C}}

\newcommand{\re}{{\rm e}}
\newcommand{\ri}{{\rm i}}
\newcommand{\rd}{{\rm d}}

\newcommand{\mU}{\mathsf{U}}

\newcommand{\mx}{\mathsf{x}}

\newcommand{\mm}{\mathsf{p}}

\newcommand{\mH}{\mathsf{H}}
\newcommand{\mJ}{\mathsf{J}}

\newcommand{\mq}{\mathsf{q}}

\newcommand{\bs}{\boldsymbol}
\newcommand{\balpha}{\boldsymbol{\alpha}}
\newcommand{\blam}{\boldsymbol{\lambda}}
\newcommand{\bn}{\boldsymbol{n}}
\newcommand{\bse}{\boldsymbol{e}}

\newdimen\tableauside\tableauside=1.0ex
\newdimen\tableaurule\tableaurule=0.4pt
\newdimen\tableaustep
\def\phantomhrule#1{\hbox{\vbox to0pt{\hrule height\tableaurule width#1\vss}}}
\def\phantomvrule#1{\vbox{\hbox to0pt{\vrule width\tableaurule height#1\hss}}}
\def\sqr{\vbox{%
 \phantomhrule\tableaustep
 \hbox{\phantomvrule\tableaustep\kern\tableaustep\phantomvrule\tableaustep}%
 \hbox{\vbox{\phantomhrule\tableauside}\kern-\tableaurule}}}
\def\squares#1{\hbox{\count0=#1\noindent\loop\sqr
 \advance\count0 by-1 \ifnum\count0>0\repeat}}
\def\tableau#1{\vcenter{\offinterlineskip
 \tableaustep=\tableauside\advance\tableaustep by-\tableaurule
 \kern\normallineskip\hbox
 {\kern\normallineskip\vbox
 {\gettableau#1 0 }%
 \kern\normallineskip\kern\tableaurule}%
 \kern\normallineskip\kern\tableaurule}}
\def\gettableau#1{\ifnum#1=0\let\next=\null\else
\squares{#1}\let\next=\gettableau\fi\next}

\begin{document}

\allowdisplaybreaks

\newcommand{\arXivNumber}{1806.01407}

\renewcommand{\PaperNumber}{025}

\FirstPageHeading

\ShortArticleName{A Solvable Deformation of Quantum Mechanics}

\ArticleName{A Solvable Deformation of Quantum Mechanics}

\Author{Alba GRASSI~$^\dag$ and Marcos MARI\~NO~$^\ddag$}

\AuthorNameForHeading{A.~Grassi and M.~Mari\~no}

\Address{$^\dag$~Simons Center for Geometry and Physics, SUNY, Stony Brook, NY, 1194-3636, USA}
\EmailD{\href{mailto:agrassi@scgp.stonybrok.edu}{agrassi@scgp.stonybrok.edu}}

\Address{$^\ddag$~D\'epartement de Physique Th\'eorique et Section de Math\'ematiques,\\
\hphantom{$^\ddag$}~Universit\'e de Gen\`eve, Gen\`eve, CH-1211 Switzerland}
\EmailD{\href{mailto:marcos.marino@unige.ch}{marcos.marino@unige.ch}}

\ArticleDates{Received October 15, 2018, in final form March 23, 2019; Published online March 31, 2019}

\Abstract{The conventional Hamiltonian $H= p^2+ V_N(x)$, where the potential $V_N(x)$ is a polynomial of degree $N$, has been studied intensively since the birth of quantum mechanics. In some cases, its spectrum can be determined by combining the WKB method with resummation techniques. In this paper we point out that the deformed Hamiltonian $H=2 \cosh(p)+ V_N(x)$ is exactly solvable for any potential: a conjectural exact quantization condition, involving well-defined functions, can be written down in closed form, and determines the spectrum of bound states and resonances. In particular, no resummation techniques are needed. This Hamiltonian is obtained by quantizing the Seiberg--Witten curve of $\mathcal{N}=2$ Yang--Mills theory, and the exact quantization condition follows from the correspondence between spectral theory and topological strings, after taking a suitable four-dimensional limit. In this formulation, conventional quantum mechanics emerges in a scaling limit near the Argyres--Douglas superconformal point in moduli space. Although our deformed version of quantum mechanics is in many respects similar to the conventional version, it also displays new phenomena, like spontaneous parity symmetry breaking.}


\Classification{14N35; 58C40; 51P05; 81T13; 81Q60; 82B23; 81Q80}

\section{Introduction}

The search for solvable models in quantum mechanics is as old as quantum mechanics itself. A~particularly important family of models which has attracted much attention in the last century is based on Hamiltonians of the form
\begin{gather}\label{sqmh}
\mH= \mm^2 + V_N(\mx),
\end{gather}
where $V_N(x)$ is a polynomial of degree $N$. When $N=2$ (the harmonic oscillator), the spectral problem was solved in the 1920s, but for potentials of higher degree there is no obvious analytic solution. On the contrary, the Hamiltonian~(\ref{sqmh}) with $N\ge 3$ has become the textbook testing ground for approximation techniques, like stationary perturbation theory or the WKB method. Starting in the 1970s, it was realized that these approximation methods can be sometimes upgraded by using resummation techniques, providing in this way a close analogue of exact solutions. In particular, exact or uniform versions of the WKB method lead to quantization conditions for anharmonic oscillators \cite{alvarez,alvarez-casares2,bpv,ddpham,zjj3,silverstone,voros,voros-quartic,zinn-justin,zjj1,zjj2}. These conditions require Borel--\'Ecalle resummations of the WKB periods and include non-perturbative corrections to the perturbative quantization condition of Dunham \cite{dunham}, which in turn is a gene\-ra\-lization to all orders in $\hbar$ of the Bohr--Sommerfeld quantization condition.

In some cases, (\ref{sqmh}) can be solved with more powerful techniques. For example, when the potential is monic (i.e., $V_N(x)=x^N$), the ODE/IM correspondence \cite{ddt, dt} provides exact quantization conditions in
terms of non-linear integral equations of the TBA type. Voros has pointed out that the solution of~(\ref{sqmh}) arises as the fixed point of a simple iteration mechanism \cite{voros1997exact,voros-fp}, although its convergence is only guaranteed when $V_N(x)$ is monic \cite{avila}. More recently, it has been pointed out in \cite{oper} that WKB periods for arbitrary polynomial potentials can be computed exactly through a TBA system, extending in this way some of the results of \cite{ddt, dt}. The TBA system of \cite{oper} was obtained as a limiting case of the one appearing in \cite{gmn}, in the context of $\CN=2$ supersymmetric gauge theory.

These remarkable developments lead to many questions. For example, one would like to obtain explicit quantization conditions for arbitrary polynomial potentials, and to clarify the relationship between the exact WKB method and methods based on integral equations of the TBA type. Another important question is whether there exist a solvable deformation of the Hamiltonian (\ref{sqmh}). Such a deformation is likely to provide new perspectives on the original quantum-mechanical problem. In this paper, we make a step in this direction: we show that the following deformation of (\ref{sqmh}),
\begin{gather}\label{dqmh}
\mH= 2\cosh \mm+ V_N(\mx),
\end{gather}
is solvable. More precisely, exact quantization conditions determining the spectrum of~(\ref{dqmh}) can be written down {\it in closed form} for {\it any} potential $V_N(x)$, in terms of {\it actual} functions. In particular, no resummation techniques are needed.

Although the physics of this deformed version of quantum mechanics is in many respects similar to the conventional version, there are also some notable differences. We find for example that the deformed version of the symmetric double-well potential displays spontaneous parity symmetry breaking for special values of the parameters. This is of course forbidden in conventional quantum mechanics\footnote{There are some peculiar quantum mechanical models with periodic potentials where spontaneous parity symmetry breaking is possible, see for instance~\cite{1202}. However this is a very different setup from the one considered in this paper.}, and parity restoration is a quintessential example of an instanton effect\footnote{We would like to warn the reader that in this work ``instantons" refers to both quantum mechanical instantons and gauge theory instantons. We hope this will not lead to confusion.}. In our deformed version of quantum mechanics, in contrast, tunneling can be suppressed at special points in moduli space, and this makes it possible to have localized states. The suppression of tunneling also leads to another surprising phenomenon, namely, the appearance of bound states in unstable potentials, like the deformed version of the cubic oscillator.

The Hamiltonian (\ref{dqmh}) has appeared before in the literature in a somewhat disguised form. As we will discuss in detail in the next section, its eigenvalue problem leads to a difference equation which is very similar to the Baxter equation of the Toda lattice. This equation was first studied in \cite{gp}, building on previous work in \cite{gutz1,gutz2,sklyanin}. However, the work \cite{gp} did not analyze this difference equation as a conventional spectral problem, but rather as a tool to understand the spectrum of the quantum Toda lattice. As a consequence, the eigenfunctions obtained in \cite{gp} are very special, and they do not correspond to generic square-integrable functions (or to generic resonant states when~$N$ is odd). In fact, these eigenfunctions only exist for a discrete set of values of the coefficients of the potential $V_N(x)$, and the results of \cite{gp}, as they stand, do not provide information on the spectrum of~(\ref{dqmh}).

The Hamiltonian (\ref{dqmh}) has another incarnation, as the quantization of the Seiberg--Witten (SW) curve~\cite{sw} describing $\CN=2$ Yang--Mills theory with gauge group ${\rm SU}(N)$ \cite{af,klt, klty}. It is indeed well-known that this curve agrees with the spectral curve of the periodic Toda lattice \cite{rus,mar-war}. The quantization of the SW curve has been considered in the literature, triggered by work of Nekrasov--Shatashivili (NS) in~\cite{ns}. For example, formal WKB periods associated to this curve have been studied in \cite{fm-omega, fm-omega2, mirmor, mirmor2,Poghossian:2010pn}. However, and somewhat surprisingly, the actual spectral problem associated to (\ref{dqmh}) has never been analyzed (except in the case of $N=2$, where the problem reduces to the well-known (modified) Mathieu equation). The Bethe/gauge correspondence of \cite{ns} has been a powerful tool to solve the quantum Toda lattice, and it has shed new light~\cite{kt2} on the solution to the Baxter equation in~\cite{gp}, but as far as we know it does not provide tools to analyze the more generic spectral problem of~(\ref{dqmh}).

In \cite{cgm, ghm}, a quantization scheme for mirror curves to toric Calabi--Yau (CY) manifolds was proposed, and exact quantization conditions for the resulting spectral problem were conjectured. We will refer to this proposal as the TS/ST correspondence. The TS/ST correspondence has also consequences for integrable systems, since quantum mirror curves can be regarded as Baxter equations for quantum cluster integrable systems~\cite{gk}, and quantization conditions for these integrable systems can be obtained as refinements of the quantization conditions for the curve \cite{fhm,ggu,hm,huang-blowup, mz-wv2,swh,wzh}. However, as emphasized in \cite{cgm}, the TS/ST correspondence focuses on the spectral problem associated to the curve itself. This suggests that it is the appropriate framework to obtain exact quantization conditions for the Hamiltonian~(\ref{dqmh}).

To achieve this goal, we consider the geometric engineering of ${\rm SU}(N)$ SW theory \cite{kkv, selfdual}. This requires taking a particular limit of topological string theory on the appropriate CY geometry, which we will refer to as the 4d limit. It turns out that this limit can be implemented at the quantum level, and leads to many simplifications of the formalism developed in~\cite{cgm, ghm}. For example, in this limit, the quantum theta function of \cite{cgm, ghm} becomes a finite sum of terms\footnote{The 4d limit studied in this paper is different from the one considered in \cite{bgt,bgt2} where the 4d quantum theta function still involves an infinite sum of terms.}. We obtain in this way a (conjectural) exact quantization condition for~(\ref{dqmh}), which looks very much like the quantization conditions obtained with the exact WKB method in ordinary quantum mechanics:
it involves the vanishing of a sum of Voros multipliers, i.e., of exponentials of quantum WKB periods, exactly as in \cite{ddpham, voros}. An important difference with conventional quantum mechanics is that, in our case,
the quantum WKB periods are not formal power series, but actual functions, defined by convergent series in gauge theory instanton calculus~\cite{n} or by solutions of non-linear integral equations of the TBA type \cite{ns}. Therefore, our solution to~(\ref{dqmh}) involves in a crucial way the connection to $\CN=2$ supersymmetric gauge theory, albeit it is very different from the one considered in~\cite{oper} for the conventional Hamiltonian~(\ref{sqmh}).

How do we recover in our formulation the standard Hamiltonian (\ref{sqmh})? It turns out that we have to consider a~scaling limit in which the Planck constant goes to zero, and the parameters specifying the potential are zoomed in near the Argyres--Douglas (AD) superconformal point of the ${\rm SU}(N)$ SW moduli space. It has already been observed in \cite{gg-qm} that the SW curve near the AD points of ${\rm SU}(3)$ and ${\rm SU}(4)$ leads to the cubic and quartic potentials in (\ref{sqmh}), respectively (this clarifies why the quantum periods for these potentials can be obtained from the refined holomorphic anomaly, as found in \cite{cm-ha,coms,fkn}). Exact quantization conditions for ordinary quantum mechanics can then be obtained, in principle, by considering the quantization conditions for our deformed theory and taking the scaling limit. Unfortunately, this is not straightforward, since the AD points are natural boundaries for the convergence of the series obtained with gauge theory instanton calculus. In addition, straightforward iteration procedures to solve the TBA system seem to break down in the vicinity of these points. Our work sheds however a new light on the long-standing problem of finding exact quantization conditions for anharmonic oscillators: this problem can be solved by understanding the scaling regime of the NS free energy near the AD points. In particular, we expect that such a scaling regime of the TBA equations of~\cite{ns} will make contact with the constructions in~\cite{cec-delz, ddt,dt,oper}.

This paper is organized as follows. In Section~\ref{sec:sp} we set up the spectral problem that we want to solve. In particular, when $N$ is odd we have to consider resonant states. We also explain the relation between the spectrum we want to find, and the spectrum of the Toda lattice. In Section~\ref{sec-qp} we review in some detail the ingredients entering our exact quantization condition: the quantum periods associated to the SW curve, and their resummation in terms of instanton calculus and TBA-type equations. In Section \ref{sec:eqc} we present the quantization condition and we give numerical evidence that it captures correctly the solution of the spectral problem. In Section \ref{derivation} we summarize the TS/ST correspondence, and we derive from it the exact quantization condition of Section~\ref{sec:eqc}. In Section \ref{sec-qmrel} we explain how to recover the standard quantum-mechanical Hamiltonian~(\ref{sqmh}) from our deformed problem, and the connection to AD points. Finally, Section~\ref{sec-con} contains our conclusions and many open problems for the future.

\section{The spectral problem}\label{sec:sp}

\subsection{General aspects}

As we mentioned in the Introduction, the spectral problem we will consider arises in the quantization of an algebraic curve. This curve is of the form
\begin{gather}\label{curve}
\Lambda^N\big(\re^p + \re^{-p} \big)+ W_N(x)=0,
\end{gather}
where
\begin{gather}\label{potW}
W_N(x)= \sum_{k =0}^N (-1)^k x^{N-k} h_k.
\end{gather}
We will set $h_1=0$ without loss of generality. The curve (\ref{curve}) has various incarnations in mathematical physics. It is the spectral curve of the Toda lattice~\cite{bbt}, and it is the SW curve for $\CN=2$ Yang--Mills theory with gauge group ${\rm SU}(N)$ \cite{af,klt,klty, sw}. In the context of SW theory, this curve is often written as
\begin{gather}\label{swcurve}
y^2 = W_N^2(x)-4 \Lambda^{2 N},
\end{gather}
after the change of variables
\begin{gather*}
\re^p={\Lambda^{-N}\over 2}(y-W_N(x)).
\end{gather*}
The possible values of the coefficients $h_2, \dots, h_{N-1}$ form an $(N-1)$-dimensional moduli space~$\CM$, which is of paramount importance in SW theory. This moduli space is often parametrized by the variables $\zeta_I$, $I=1, \dots, N$, which are defined by
\begin{gather} \label{wnx}
W_N(x)= \prod_{i=1}^{N} ( x- \zeta_I ),
 \end{gather}
and satisfy the constraint
\begin{gather}\label{al-cons}
\sum_{I=1}^N \zeta_I=0.
\end{gather}
The coefficients $h_k$ are given by the elementary symmetric functions of the moduli $\zeta_I$. In terms of Schur polynomials $s_R(\zeta_I)$, labelled by Young tableaux~$R$, one has~\cite{macdonald}
\begin{gather*}
 h_k =s_{(1^k)} (\zeta_I ),
\end{gather*}
where $(1^k)$ is the Young tableau with $k$ boxes and $k$ rows. We note that the permutation group~$S_N$ acts on the moduli~$\zeta_I$ by permutation of their labels. This action leaves the~$h_k$ invariant, therefore the correspondence between the $h_k$ and the $\zeta_I$ is multivalued: two points~$\zeta_I$,~$\zeta'_I$ related by a permutation correspond to the same point in moduli space. We will usually denote
\begin{gather*}
h=-h_2.
\end{gather*}
We will also use sometimes the terminology of integrable systems and call the $h_k$ Hamiltonians, since they correspond to the Hamiltonians of the Toda lattice.

The quantization of (\ref{curve}) is done by promoting $\mx$, $\mm$ to Heisenberg operators satisfying the canonical commutation relation
\begin{gather}\label{4dpc}
[\mx, \mm]=\ri \hbar.
\end{gather}
We will consider the following spectral problem, associated to the quantization of the spectral curve (\ref{curve}):
\begin{gather}\label{specp1}
\big\{ \Lambda^N\big( \re^\mm + \re^{-\mm} \big)+ W_N(\mx)\big\} \big|\psi \rangle=0,
\end{gather}
which can be written, in the position representation, as a difference equation for the wave\-func\-tion~$\psi(x)$:
\begin{gather}\label{baxter}
\Lambda^N\big( \psi(x+\ri \hbar) + \psi(x-\ri \hbar) \big)+ W_N(x) \psi(x)=0.
\end{gather}
Equivalently, we can write (\ref{specp1}) as
\begin{gather}\label{specp2}
\mH_N |\psi \rangle= (-1)^{N-1} h_N| \psi \rangle,
\end{gather}
where the Hamiltonian is given by
\begin{gather}\label{q-ham}
\mH_N=\Lambda^N\big( \re^\mm+\re^{-\mm} \big) +V_N(\mx),
\end{gather}
and
\begin{gather*}
V_N(\mx)= \sum_{k=0}^{N-1} (-1)^k \mx^{N-k} h_k
\end{gather*}
is a degree $N$ potential. The coefficient $(-1)^{N-1} h_N$ appearing in~(\ref{potW}) is then interpreted as the eigenvalue of the Hamiltonian (this is the point of view used in~\cite{cgm, ghm} in the quantization of mirror curves). We note that, when $N=2$, the Hamiltonian is
\begin{gather*}
\mH_2= 2 \Lambda^2 \cosh(\mm)+ \mx^2,
\end{gather*}
and it is equivalent, after a canonical transformation, to the Schr\"odinger operator appearing in the modified Mathieu equation. For $N\ge 3$, $\mH_N$ is a deformation of the standard Hamiltonian for a non-relativistic particle in an arbitrary potential, in one dimension:
\begin{gather*}
\mH^{\rm QM}_N=\Lambda^N \mm^2 + V_N(\mx).
\end{gather*}
The relation to the standard quantum mechanical spectral problem will be made more precise in Section~\ref{sec-qmrel}.

Let us now examine in detail the spectral problem (\ref{specp2}). We will assume for simplicity that the coefficients in the Hamiltonian $\Lambda, h_2, \dots, h_{N-2}$ are real\footnote{This condition can be easily relaxed and the generalizations will be studied elsewhere.}. Not surprisingly, the spectrum depends crucially on the parity of~$N$. When $N$ is even, the potential~$V_N(x)$ is confining, and~(\ref{specp2}) leads to an infinite, discrete spectrum as shown in \cite[Proposition~2.1]{lst}. A heuristic argument supporting this conclusion is that the region in phase space
\begin{gather}\label{cre}
\CR(E)= \big\{ (x, p) \in \IR^2\colon 2 \cosh(p) + V_N(x) \le E\big\}
\end{gather}
is compact and has finite volume, as measured by the canonical symplectic form $\omega=\rd x \wedge \rd p$. In fact, the spectrum should be captured at large quantum numbers by the Bohr--Sommerfeld quantization condition
\begin{gather*}
{\rm vol}_0(E) \approx 2 \pi \hbar \left(n+{1\over 2}\right),
\end{gather*}
where ${\rm vol}_0(E)$ is the volume of the region (\ref{cre}).

When $N$ is odd, the situation is very different, since the potential $V_N(x)$ is unbounded from below and there are no bound states. However, as in the case of anharmonic oscillators of odd degree in standard quantum mechanics \cite{calicetiodd, kp}, we expect to have an infinite tower of {\it resonances}, associated to the so-called Gamow states. We recall that Gamow states are described by wavefunctions $\psi(x)$ which satisfy appropriate boundary conditions. In the case of odd potentials, one requires decay at infinity in the direction in which the potential grows, and Gamow--Siegert boundary conditions (i.e., purely outgoing states) in the direction in which the potential decreases. In this paper we will follow a down-to-earth approach to the spectral problem~(\ref{specp2}) when $N$ is odd, and we will define and compute the resonant eigenvalues by using the complex dilatation method~\cite{bender-resonance}. We recall that, in this method, one considers the group of complex dilatations, acting on the Heisenberg operators $\mx$, $\mm$ as
\begin{gather*}
\mU_{ \ri \theta} \mx \mU_{\ri \theta}^{-1} = \re^{\ri \theta}\mx, \qquad \mU_{\ri \theta} \mm \mU^{-1}_{\ri \theta} = \re^{-\ri \theta} \mm.
\end{gather*}
We usually take $\theta$ to be real, and we consider the rotated Hamiltonian
\begin{gather}\label{Htheta}
\mH_N (\theta)= \mU_{ \ri \theta} \mH_N \mU^{-1}_{\ri \theta} =\exp\big( \re^{-\ri \theta} \mm\big)+ \exp\big({-}\re^{-\ri \theta} \mm\big)+ V_N\big(\re^{\ri \theta} \mx\big).
\end{gather}
When $\theta\not=0$, his Hamiltonian has square-integrable eigenfunctions with complex eigenvalues, which are independent of the precise value of~$\theta$ beyond a given threshold, as explained in~\cite{bender-resonance}. These eigenfunctions are obtained by acting with $\mU_{\ri \theta}$ on Gamow states.

\begin{figure} \centering
\includegraphics[scale=0.45]{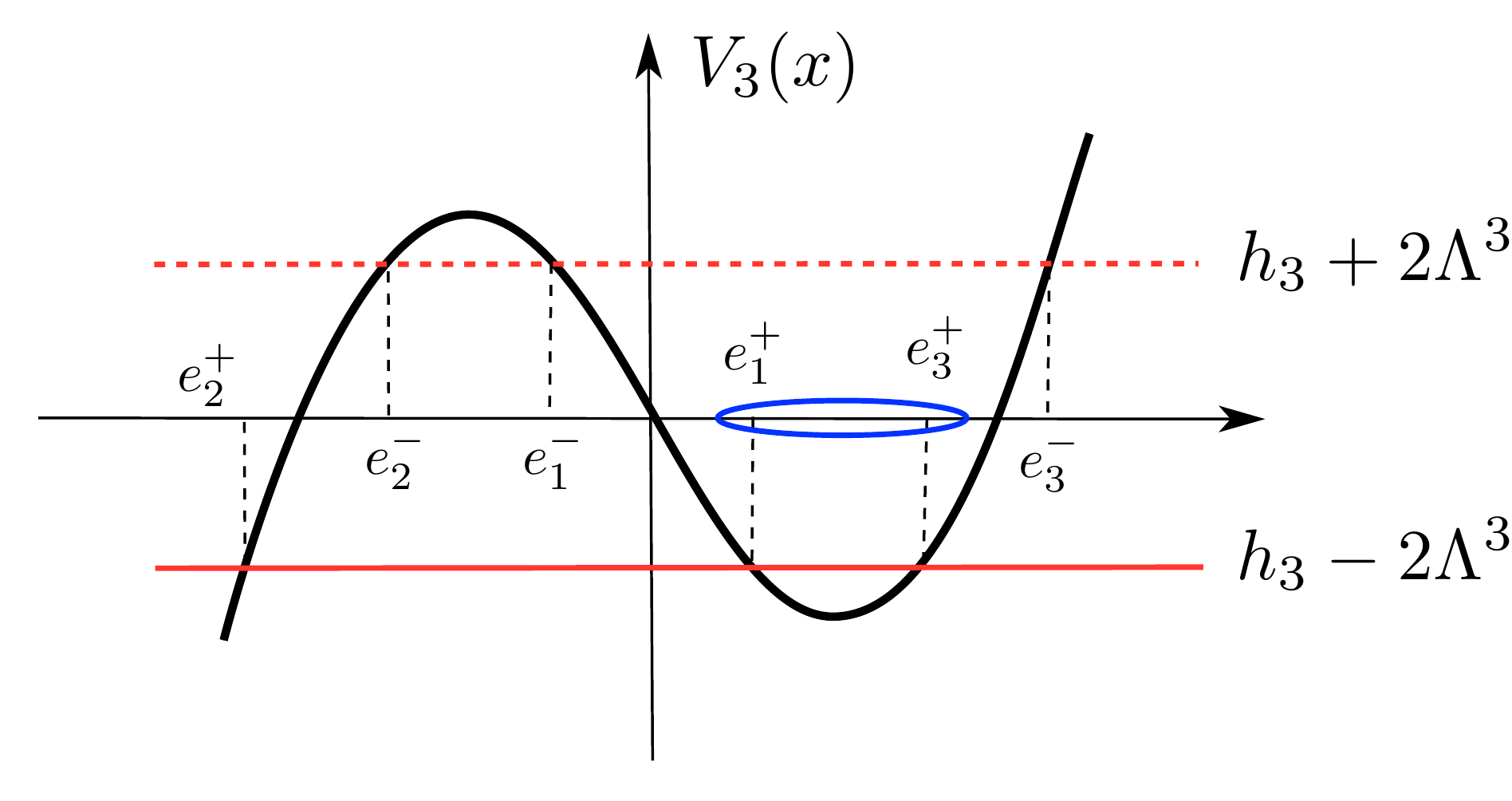}
\caption{The potential $V_3(x)$ and the branch points (\ref{swroots}). Classical motion is allowed in the compact interval $[e_1^+, e_3^+]$ and in the semi-infinite interval $(-\infty, e_2^+]$.} \label{su3fig}
\end{figure}

It is illuminating to consider the {\it classical} dynamics associated to the classical counterpart of the Hamiltonian (\ref{q-ham}), where $(-1)^{N-1} h_N=E$ can be now regarded as the conserved energy of the system. Since $\cosh(p) \ge 1$ for real $p$, the classical motion is confined to the intervals where
\begin{gather*}
W_N(x) \ge -2 \Lambda^N.
\end{gather*}
In these intervals, the EOM is given by
\begin{gather}\label{xeom}
\dot x= {\sqrt{W^2_N(x) - 4 \Lambda^{2N}}}.
\end{gather}
The function inside the square root (\ref{xeom}) is precisely the SW curve (\ref{swcurve}). If we write $W_N(x)$ as in (\ref{wnx}), where $\zeta_I=\zeta_I(h_2, \dots, h_N)$, the zeroes of this function (which are the branch points of the hyperelliptic curve~(\ref{swcurve})) are given by~\cite{klt}
\begin{gather}\label{swroots}
e^\pm_I (h_2, \dots, h_N, \Lambda)= \zeta_I\big(h_2, \dots, h_N\pm 2 \Lambda^N \big).
\end{gather}
In Fig.~\ref{su3fig} we show the potential $V_3(x)$ and the points $e_I^\pm$, $I=1, 2,3$. There is one compact interval where motion is allowed, namely $[e_1^+, e_3^+]$. This is one of the intervals of instability of the Toda lattice (see, e.g.,~\cite{bbt}). The other interval of instability is $[e_2^-, e_1^-]$, but in our problem motion is not allowed in this interval, since~$p$ is complex there and of the form ${\rm Re}(p)+ \ri \pi$. As we will see, however, this interval, as well as the intervals between the points~$e_I^+$ and~$e^-_I$, play a~r\^ole in the quantum theory.

\subsection{Relation to the Toda lattice}

The difference equation (\ref{baxter}) is very similar to the Baxter equation of the quantum-mechanical Toda lattice \cite{gutz1, gutz2,kl, gp, sklyanin}. In solving this equation, one imposes very stringent boundary conditions on $\psi(x)$, which are only satisfied for discrete values of the $h_k$, $k=2, \dots, N-1$. These values lead to an infinite set of points in moduli space which give the spectrum of the commuting Hamiltonians of the underlying integrable system. We will call these points, by a~slight abuse of language, the {\it Toda lattice points}.

In contrast, in the spectral problem we are considering, we impose conventional boundary conditions for $\psi(x)$: square integrability, when $N$ is even, and Gamow--Siegert boundary conditions
when $N$ is odd. With these conventional boundary condition, (\ref{specp2}) leads to an infinite tower of discrete eigenvalues $h^{(n)}_N (h_2, \dots, h_{N-1})$, $n=0,1,2, \dots$, as a function of the parameters $h_2, \dots, h_{N-1}$. Geometrically, the allowed values of the moduli form a discrete family of subma\-ni\-folds~$\CS_n$ of codimension one inside the moduli space, labelled by the non-negative integer~$n$~\cite{cgm}.

It is easy to show that the Toda lattice points belong to the union of submanifolds $\cup_{n=0}^\infty \CS_n$. To see this, we recall that the Baxter equation for the Toda lattice is~\cite{gp}
\begin{gather*}
 ( \ri \Lambda )^N Q(x+\ri \hbar) + ( -\ri \Lambda )^N Q(x-\ri \hbar) = W_N(x) Q(x).
\end{gather*}
The solutions to this equation considered in \cite{gp} satisfy two conditions: they are entire in the complex $x$-plane, and they have the following decay behavior at infinity along the real axis,
\begin{gather*}
Q(x) \approx \re^{- {N \pi \over 2 \hbar} |x|}, \qquad x\rightarrow \pm \infty.
\end{gather*}
These conditions restrict the values of $h_k$, $k=2, \dots, N$ to a discrete set of values. Let us now consider the following function,
\begin{gather*}
\varphi(x) = \re^{ {\pi (N-2) x \over 2 \hbar}} Q(x).
\end{gather*}
It is immediate to check that $\varphi(x)$ is entire, satisfies the difference equation (\ref{baxter}), and, for $N>1$, it decreases exponentially at infinity on the real line. In particular, $\varphi(x) \in L^2(\IR)$. Therefore, when $N$ is even, it is an admissible eigenfunction of~(\ref{specp2}). When $N$ is odd, it is an admissible Gamow state, but a~very special one, since its resonant energy is real and it is described by a~square integrable wavefunction. We conclude that, if $h_2, \dots, h_N$ lead to a solution of the Baxter equation, $h_N$ is an eigenvalue or resonance of (\ref{specp2}) for the values of the parameters given by $h_2, \dots, h_{N-1}$.

\subsection{Examples and new phenomena}

Although the deformation of quantum mechanics we are studying is similar to standard one-dimensional quantum mechanics in many respects, it also displays some striking differences. For example, in the deformed version of the symmetric double-well potential, there are degenerate states and parity symmetry breaking at Toda lattice points. These are forbidden in conventional quantum mechanics, yet they occur here. In the deformed version of the cubic oscillator, we find special resonances with purely real eigenvalues. We note that restoration of parity symmetry and complex eigenvalues are the signal of non-perturbative tunneling effects in the double-well and the cubic oscillator, respectively. It turns out that, in our deformed quantum mechanics, tunneling effects are suppressed at the Toda lattice points, leading to this unconventional phenomena. Let us now illustrate these general considerations with a closer look at the spectral problem~(\ref{specp2}) for $N=3$, $N=4$.

\begin{figure}[tb]\centering
\begin{tabular}{cc}
\resizebox{70mm}{!}{\includegraphics{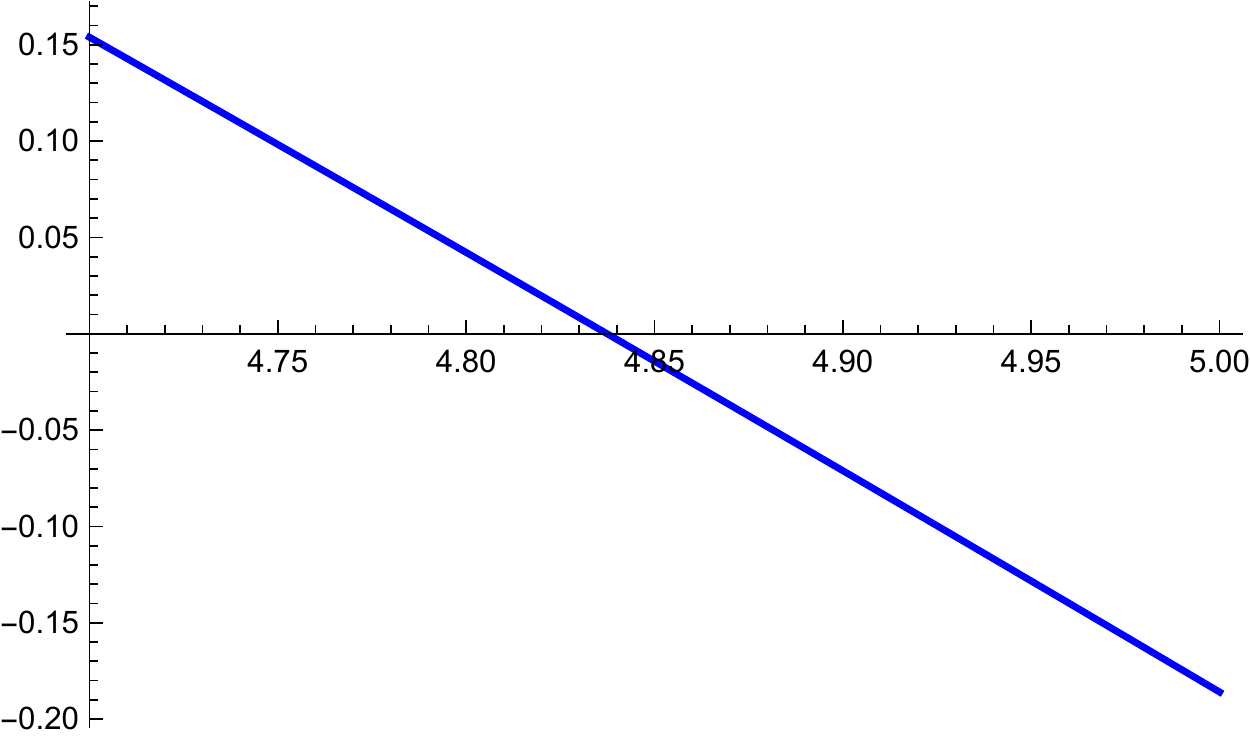}}
\hspace{3mm}
&
\resizebox{65mm}{!}{\includegraphics{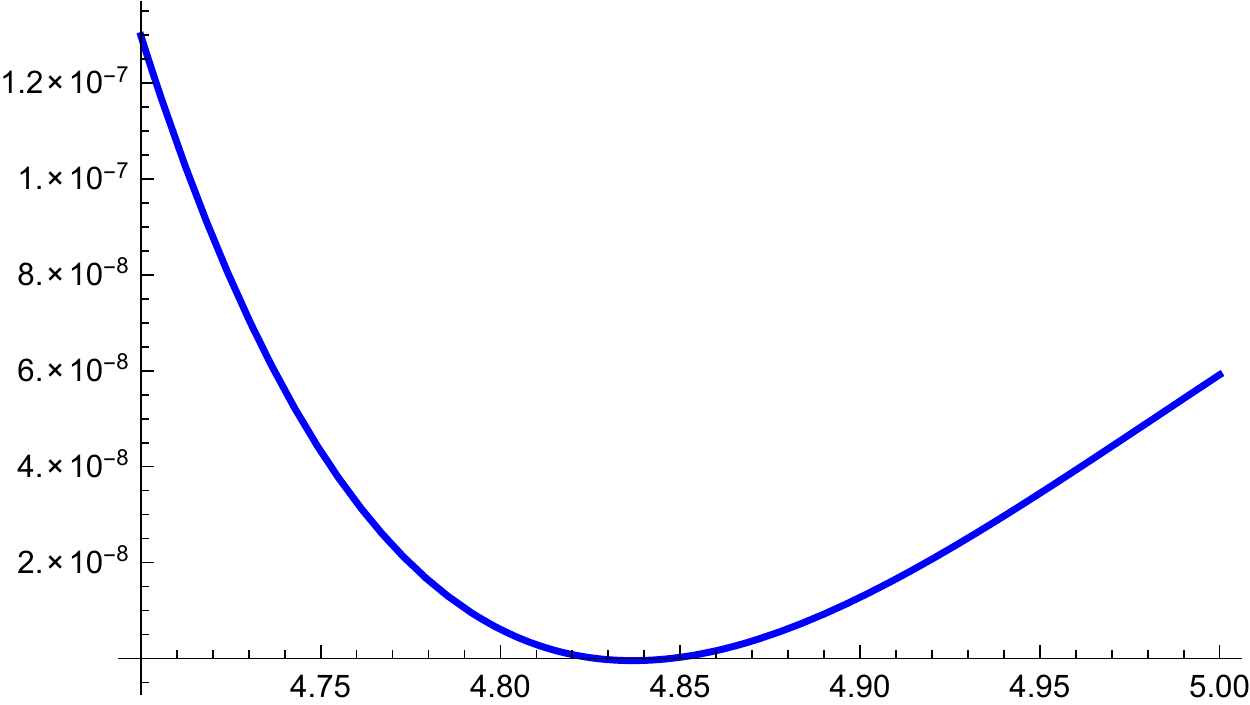}}
\end{tabular}
\caption{${\rm Re}\big(h^{(0)}_3\big)$ (left) and ${\rm Im}\big(h^{(0)}_3\big)$ (right) as a function of $h$, for $\hbar=\Lambda=1$, computed by using complex dilatation techniques. When $h=4.837{\dots}$, corresponding to the first eigenvalue of the first Hamiltonian of the ${\rm SU}(3)$ Toda lattice, one finds that $h_3^{(0)}=0$. In particular, its imaginary part vanishes.}\label{toda3-res}
\end{figure}

When $N=3$, the Hamiltonian is
\begin{gather}\label{n3hamil}
\mH_3 = \Lambda^3\big( \re^{\mm}+ \re^{-\mm} \big)+\mx^3 - h \mx.
\end{gather}
This can be regarded as a deformation of the cubic anharmonic oscillator. When $h >0$, and in order to calculate its resonant eigenvalues, it is useful to write the potential term
\begin{gather*}
V_3(x) = x^3-h x
\end{gather*}
as a perturbed harmonic oscillator, by expanding around the minimum at
\begin{gather*}
x_0= {\sqrt{h \over 3}}.
\end{gather*}
We find
\begin{gather*}
V_3(x)= -2 \left({h \over 3} \right)^{3/2} + q^2 +2g q^3,
\end{gather*}
where
\begin{gather*}
q=(3 h)^{1/4} (x-x_0), \qquad g={1\over 2 (3 h)^{3/4}}.
\end{gather*}
We are let to consider the Hamiltonian,
\begin{gather*}
\widetilde {\mH}=\Lambda^3 \cosh(\mm)+{\mq^2 \over 2} + g\mq^3,
\end{gather*}
where $\mq$, $\mm$ satisfy the commutation relations
\begin{gather*}
[\mq, \mm]= \ri (3 h)^{1/4} \hbar.
\end{gather*}
Let us denote by $\lambda_n$ the (resonant) eigenvalues of $\widetilde{\mH}$, calculated by complex dilatation techniques. The eigenvalue $h_3$ in~(\ref{specp2}) is then given by
\begin{gather*}
h^{(n)}_3= 2 \lambda_n- 2\left({h \over 3} \right)^{3/2}, \qquad n=0,1, 2, \dots.
\end{gather*}

In Fig.~\ref{toda3-res} we show the real and the imaginary part of the eigenvalue $h^{(0)}_3$ of $\mH_3$ for the ground state, as a function of $h$, and for $\Lambda=\hbar=1$. In general, as expected with resonances, we find a complex eigenvalue, with a small imaginary part\footnote{Resonant energies appear in pairs, related by complex conjugation. In this paper we always choose a positive sign for the imaginary part of the energy, both in numerical calculations and in the quantization conditions.}. However, when $h=4.837{\dots}$ is the first eigenvalue of the first Hamiltonian of the ${\rm SU}(3)$ Toda lattice \cite{matsuyama}, the imaginary part of $h_3^{(0)}$ {\it vanishes} (its real part vanishes too, in this case). As a consequence, the Toda lattice point $(h, h_3)= (4.837{\dots}, 0)$, which is the ground state of the ${\rm SU}(3)$ Toda lattice for $\hbar=1$, belongs to the resonant spectrum of the Hamiltonian $\mH_3$, in accord with our general discussion. However, it leads to a state with a {\it real} eigenvalue\footnote{Similar special resonances were observed in \cite{cgum} for quantized mirror curves.}, which should be then regarded as a bound state, rather than as a resonant state.

\begin{figure}[tb]\centering
\includegraphics[scale=0.45]{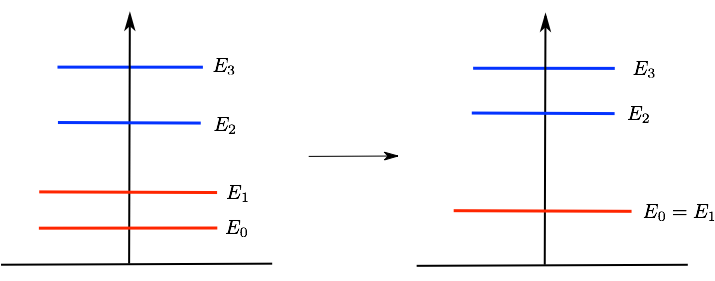}
\caption{Parity symmetry breaking in deformed quantum mechanics. In the deformed, symmetric double-well, as the parameter $h$ approaches the values corresponding to Toda lattice points with quantum numbers $(0,n,0)$, the ground state and the first excited state, which have opposite parity, become degenerate.}\label{psbfig}
\end{figure}

\begin{figure}[tb]\centering
\includegraphics[scale=0.7]{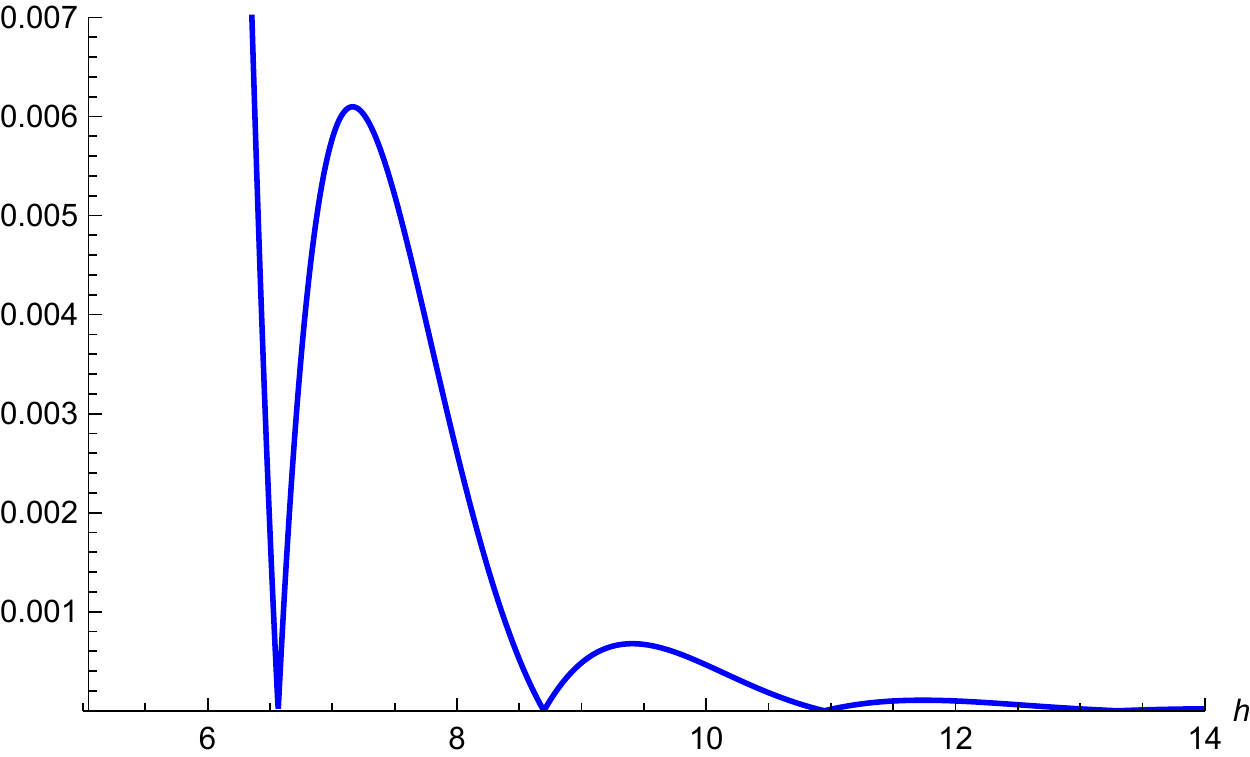}

 \caption{The difference $E_1-E_0$ between the first two energy levels of \eqref{H40} as a function of $h$. The zeroes are located at the Toda lattice points where $h$ takes the values $h^{ \rm T }(0,n,0)$ for $n=0,1,2, 3$. We set $\Lambda=\hbar=1$. }\label{e0e1full}
\end{figure}

Let us now consider the deformed version of the symmetric double-well potential. This is the Hamiltonian (\ref{dqmh}) with $N=4$ and $h_3=0$:
\begin{gather} \label{H40} \mH_4 = \Lambda^4 \left( \re^{\mm}+ \re^{-\mm} \right)+\mx^4 - h \mx^2 .\end{gather}
We denote by
\begin{gather}\label{eigH40}
E_n(h)=-h_4^{(n)}, \qquad n=0,1,2, \dots
\end{gather}
the eigenvalues of \eqref{H40}. Since the Hamiltonian~(\ref{H40}) is symmetric, the eigenstates have a~definite parity. As we explained above, the Toda lattice points belong to the spectrum of this Hamiltonian. In the case of $N=4$, these points can be labelled by three quantum numbers inherited from the Einstein--Brillouin--Keller quantization conditions. By following the conventions of~\cite{matsuyama} we denote them as
\begin{gather}\label{4qn}
 \{ \ell_1, \ell_2, \ell_3 \}, \qquad \ell_i =0,1,2, \dots. \end{gather}
The corresponding values of $h$ at these points will be denoted by
\begin{gather*}
h^{\rm T}( \ell_1, \ell_2, \ell_3).
\end{gather*}
When $h_3=0$ however one has $\ell_3=\ell_1$ and the Toda lattice points are labelled as
\begin{gather*} \{ \ell_1, \ell_2, \ell_1 \}. \end{gather*}
We have for instance
\begin{gather*} h^{\rm T}(0,0,0)=6.562{\dots}, \qquad h^{ \rm T}(1,0,1)=9.636{\dots}.\end{gather*}
It turns out that, at the Toda lattice points, there are two successive eigenvalues of \eqref{eigH40} which become identical, and we find then two degenerate states of opposite parity. More precisely, we have
\begin{gather} \label{degeq}
E_{2k}\left( h^{ \rm T } (k, n ,k) \right)=E_{2k+1}\left( h^{ \rm T } (k, n ,k) \right), \qquad n=0,1,2, \dots.
\end{gather}
In particular, when $k=0$, the ground state becomes degenerate and we have parity-symmetry breaking. This is illustrated schematically in Fig.~\ref{psbfig}. A more precise description of this phenomenon is given in Fig.~\ref{e0e1full}, where we plot the difference $E_1-E_0$ between the first excited state energy and the ground state energy, as a function of~$h$. At the values $h= h^{\rm T}(0,n,0)$ (the first four of them are shown in the figure), the energies are equal and the ground state is degenerate.

It is well-known that parity symmetry breaking, and more generally degenerate bound states, cannot occur for standard Schr\"odinger operators (see for example~\cite{qfsm, ll}). This relies heavily on having a kinetic term of the form $\mm^2$, therefore the standard arguments do not apply in our case. In the case of the conventional, symmetric double-well potential, the absence of degenerate states for low-lying levels is due to tunneling through the barrier, and the energy difference between successive low-lying levels of opposite parity is an instanton effect (see, e.g., \cite{mmbook, zjj1}). Therefore, the degeneracy (\ref{degeq}) indicates that, for certain pairs
of successive levels, these non-perturbative effects disappear at the Toda lattice points. As we will see in Section~\ref{sec:eqc}, our exact quantization condition gives a quantitative explanation of this phenomenon.

We should mention that the deformed, symmetric double well (\ref{H40}) leads to other significant differences with standard quantum mechanics properties. For example, for some values of $h$, the eigenfunction representing the ground state might be odd, or might change sign.

\section{The quantum periods}\label{sec-qp}
As in the conventional exact WKB method, the basic ingredients in our exact quantization conditions are the quantum periods associated to the curve (\ref{curve}). This curve has genus $N-1$ and it comes equipped with $2(N-1)$ one-cycles. We can choose an appropriate symplectic homology basis for these cycles, which we will denote by $\{A_i, B_i\}_{i=1, \dots, N-1}$. In practice, such a~basis is obtained by considering appropriate contours around the branch points (\ref{swroots}). Let us now introduce the Liouville form
\begin{gather*}
\lambda= p(x) \rd x . \end{gather*}
The periods of this form on the symplectic basis of cycles will be denoted by
\begin{gather*}
a_i^{(0)} = \oint_{A_i} \lambda, \qquad {\partial F_0 \over \partial a_i} = \oint_{B_i} \lambda,
\end{gather*}
where $i=1, \dots, N-1$. These are the periods appearing in SW theory. The cycles $A_i$, $B_i$ are the ones associated to the so-called electric frame, and they are appropriate for the semiclassical regime in which the moduli are large (see, e.g.,~\cite{klt} for more details). These periods give the exact masses of BPS states in the underlying $\CN=2$ Yang--Mills theory, and they define the SW prepotential $F_0(a_1, \dots, a_{N-1})$.

Let us now use an all-orders WKB ansatz to solve the difference equation (\ref{baxter}), as in \cite{dingle,dunham,gp},
\begin{gather*}
\psi(x) =\exp \left( {\ri \over \hbar} \sum_{n\ge 0} S_n(x) \hbar^n \right).
\end{gather*}
Then, if we write
\begin{gather*}
S_n(x)= \int^x Y_n(x) \rd x,
\end{gather*}
we have that $Y_0(x)= p(x)$. The differential
\begin{gather*}
\lambda(\hbar)= \sum_{\ell\ge 0} Y_{2\ell}(x) \hbar^{2 \ell},
\end{gather*}
which can be regarded as a deformation of the standard Liouville form, leads to the so-called quantum $A$- and $B$-periods
\begin{gather}
a_i= \oint_{A_i} \lambda(\hbar)= \sum_{\ell \ge0} a_i^{(\ell)} \hbar^{2 \ell},\nonumber\\
{\partial F_{\rm NS} \over \partial a_i}= \oint_{B_i} \lambda(\hbar)= \sum_{\ell \ge0} {\partial F^{(\ell)}_{\rm NS} \over \partial a_i}\hbar^{2 \ell},\label{qpers-int}
\end{gather}
which are {\it a priori} formal power series in $\hbar$, depending on the moduli of the curve $h_2, \dots, h_N$ (as in the standard WKB method, the functions $Y_n(x)$, for odd $n>1$, are total derivatives, and do not contribute to the period integrals). These relations define the all-orders WKB free energy $F_{\rm NS} (a_1, \dots, a_{N-1}, \hbar)$. Its leading order term in the $\hbar$ expansion, $F_{\rm NS}^{(0)}$, is the SW prepoten\-tial~$F_0$.

One way to obtain actual functions from the formal power series defining the quantum periods is to use the technology of Borel--\'Ecalle resummation, resurgence theory, and the like (see \cite{abs,mmlargen, ss} for pedagogical introductions to resurgence in general, and~\cite{bpv,dpham} for applications in quantum mechanics). It turns out that, in the case of the difference equation (\ref{baxter}) associated to the SW curve, the quantum periods can be promoted to {\it actual} functions by using instanton calculus in supersymmetric gauge theory. As argued in \cite{mirmor, mirmor2}, the all-orders WKB free energy agrees with the NS limit of the instanton free energy of $\CN=2$ supersymmetric Yang--Mills theory in four dimensions (hence our notation in the second line of~(\ref{qpers-int})). This quantity was introduced in~\cite{n,ns} (see also~\cite{bfmt,tachikawa} for pedagogical reviews). We will now summarize some basic facts about the NS free energy which will be needed in this paper. Since our quantization conditions will be obtained from the topological string theory/five dimensional point of view, we will present the results first in 5d, and then we will explain how to implement the 4d limit.

Our starting point is then the partition function of the 5d supersymmetric Yang--Mills theory with gauge group ${\rm SU}(N)$ in the Omega-background \cite{n,no2}. This partition function agrees~\cite{taki} with the partition function of the refined topological string on the~$A_N$ CY geometry which engineers the corresponding gauge theory~\cite{kkv}. This object can then be computed with instanton counting, as in \cite{n}, or with the refined topological vertex \cite{akmv,ikv}. In order to write it down we need some notation. We introduce a vector of Young tableaux or partitions $Y_I$, $I=1, \dots, N$,
\begin{gather*}
\boldsymbol{Y}=(Y_1, \dots, Y_N).
\end{gather*}
A box in a Young tableau will be denoted by $s=(i,j)$. Given a partition $Y=(y_1, y_2,\dots)$ and a box $s=(i,j)$ (not necessarily in the partition $Y$), we define
\begin{gather*}
h_Y(s)=y_i-j, \qquad v_Y(s)= y^t_j-i,
\end{gather*}
where $Y^t=\big(y_1^t, y_2^t,\dots\big)$ is the transposed partition. The total number of boxes in the Young tableau $Y$ is denoted by
\begin{gather*}
\ell(Y)= \sum_i y_i.
\end{gather*}
We also introduce the parameters $\epsilon_{1,2}$ characterizing the Omega-background through their exponents,
\begin{gather*}
q=\re^{\epsilon_1}, \qquad t=\re^{-\epsilon_2}.
\end{gather*}
We use conventions in which the standard topological string is obtained when $\epsilon_1+ \epsilon_2=0$. The NS limit~\cite{ns} corresponds to
\begin{gather}\label{nslimit}
\epsilon_1= \ri \hbar, \qquad \epsilon_2=0.
\end{gather}
The partition function of the 5d ${\rm SU}(N)$ theory, $Z_{\rm 5d}(\bs{Q}; \epsilon_1, \epsilon_2)$, depends on $\epsilon_{1,2}$ and on the exponentiated variables
\begin{gather}\label{Qal}
Q_I=\re^{\alpha_I}, \qquad I=1, \dots, N,
\end{gather}
which satisfy the constraint
\begin{gather*}
\sum_{I=1}^N \alpha_I=0.
\end{gather*}
We will also denote
\begin{gather*}
Q_{IJ}= \re^{\alpha_I-\alpha_J}.
\end{gather*}
$Z_{\rm 5d}(\bs{Q}; \epsilon_1, \epsilon_2)$ can be written as a sum over Young tableaux\footnote{We would like to warn the reader that $Z_{\rm 5d}$ denotes here only the instanton part of the ${\rm 5d}$ partition function, in particular this does not contain the perturbative (one-loop) part.},
\begin{gather*}
Z_{\rm 5d} (\bs{Q}; \epsilon_1, \epsilon_2)=\sum_{\boldsymbol{Y}}\big( (-1)^N \Lambda_{\rm 5d}^{2 N}\big)^{\ell(\boldsymbol{Y})} \CZ^{\rm 5d}_{\boldsymbol{Y}},
\end{gather*}
where
\begin{gather*}
\ell( \boldsymbol{Y})= \sum_{I=1}^N \ell(Y_I),
\end{gather*}
and
\begin{gather*}
\CZ^{\rm 5d}_{\bs{Y}}=\prod_{I,J=1}^N \prod_{s \in Y_I} {1\over 1- Q_{JI} q^{v_{Y_J}(s)} t^{h_{Y_I}(s)+1}} \prod_{s\in Y_J}
{1\over 1- Q_{JI} q^{-v_{Y_I}(s)-1} t^{-h_{Y_J}(s)}}.
\end{gather*}
We note that we have introduced an additional minus sign in the instanton counting para\-me\-ter~$\Lambda_{\rm 5d}^{2 N}$, as compared to the standard formulae for the instanton partition function. This is the correct sign in order to make contact with the spectral problems we are considering (it also appears in the formulae in~\cite{hm}). The NS free energy in 5d is then defined as the NS limit (\ref{nslimit}) of the free energy,
\begin{gather}\label{5dns}
F_{\rm NS, inst}^{\rm 5d}(\bs{Q}, \hbar) =\ri \lim_{\epsilon_2 \rightarrow 0} \epsilon_2 \log Z_{\rm 5d}(\bs{Q}; \ri \hbar , \epsilon_2).
\end{gather}

The SW curve of the 5d theory is given by \cite{nek5}
\begin{gather}\label{rt-sc}
\Lambda_{\rm 5d}^{N} \big( \re^p + \re^{-p} \big)+ W_N^{\rm 5d}(z)=0.
\end{gather}
In this equation,
\begin{gather*}
W^{\rm 5d}_N(z)=\prod_{i=1}^{N} 2 \sinh \left( {x- \zeta_I \over 2}\right)= \sum_{k=0}^N (-1)^k z^{N-2 k} H_k,
\end{gather*}
where we have denoted $z=\re^x$. The quantities $\zeta_I$ are moduli for the curve, just as in~(\ref{wnx}), and they also satisfy the constraint~(\ref{al-cons}). The $H_k$ are elementary symmetric functions of the~$\re^{\zeta_I}$, and we have $H_0=H_N=1$. The $A$-periods of this curve are functions of the moduli $H_k$, and in the presence of the Omega background they depend as well on the parameters~$\epsilon_{1,2}$. The resulting deformed $A$-periods are linear combinations of the parameters $\alpha_I$ introduced in (\ref{Qal}). Therefore, we can write,
\begin{gather}\label{qmm}
H_k= H_k (\bs{Q}; \epsilon_1, \epsilon_2), \qquad k=1, \dots, N-1.
\end{gather}
If $\epsilon_i \to 0$, and in the semiclassical region $|H_k|\gg 1$ we have that $\alpha_I \approx \zeta_I$. In the NS limit, the relation~(\ref{qmm}) is sometimes called the quantum mirror map~\cite{acdkv}. The explicit form of~(\ref{qmm}) can be obtained by using instanton calculus in gauge theory \cite{bk,bkk,francisco,fm-wilson,lmn,sciarappa1}. Let us define
\begin{gather*}
\CW= \sum_{I=1}^N \re^{\alpha_I}, \qquad
\CV_{\boldsymbol{Y}}=\sum_{I=1}^N \re^{\alpha_I}\sum_{(k,l)\in Y_I}\re^{ -(k-1) \epsilon_1 - (l-1)\epsilon_2}.
\end{gather*}
We also need the equivariant Chern character,
\begin{gather}\label{cher-char}
{\rm Ch}_{\boldsymbol{Y}}(\CE)= \CW- \big(1-\re^{-\epsilon_1}\big) \big(1-\re^{- \epsilon_2}\big) \CV_{\boldsymbol{Y}}.
\end{gather}
The vev of a 5d Wilson loop in the fundamental representation is given by
\begin{gather}\label{wl-fund}
W_{\tableau{1}}={1\over Z_{\rm 5d}} \sum_{\boldsymbol{Y}} \big((-1)^N\Lambda_{\rm 5d}^{2N} \big)^{\ell (\boldsymbol{Y})} {\rm Ch}_{\boldsymbol{Y}}(\CE) \CZ^{\rm 5d}_{\boldsymbol{Y}}.
\end{gather}
Wilson loops in higher representations, which we will denote by $W_R$, can be obtained by using the Adams operation in representation theory. Let $\bs{k}=(k_1, k_2,\dots)$ a vector of non-negative entries, and let us define
\begin{gather*}
{\rm Ch}_{\bs{k},\boldsymbol{Y}}(\CE)= \prod_{j\ge 1} \big( {\rm Ch}_{\boldsymbol{Y}}\big(\CE^j\big) \big)^{k_j},
\end{gather*}
where the power $\CE^j$ means that the variables appearing in ${\rm Ch}_{\boldsymbol{Y}}(\CE)$ are rescaled as
\begin{alignat*}{3}
& \alpha_I \rightarrow j \alpha_I, \qquad && I=1, \dots, N, & \\
& \epsilon_i \rightarrow j \epsilon_i, \qquad && i=1,2.&
\end{alignat*}
Let us regard $\bs{k}$ as a conjugacy class of the symmetric group of order
\begin{gather*}
\ell({\bs k})= \sum_{j\ge 1} j k_j,
\end{gather*}
and let us denote by $\chi_\CR(\bs{k})$ the character of $\bs{k}$ in the representation $\CR$. Then, we define,
\begin{gather*}
{\rm Ch}_{\CR,\boldsymbol{Y}}(\CE)=\sum_{\bs{k}} {\chi_\CR(\bs{k}) \over z_{\bs{k}}} {\rm Ch}_{\bs{k},\boldsymbol{Y}}(\CE),
\end{gather*}
where
\begin{gather}\label{zk}
z_{\bs k}= \prod_{j\ge 1} k_j! j^{k_j}.
\end{gather}
For example, the symmetric and antisymmetric representation are obtained as~\cite{bk}
\begin{gather*}
{\rm Ch}_{\tableau{2},\boldsymbol{Y}}(\CE) ={1\over 2} \big( ( {\rm Ch}_{\boldsymbol{Y}}(\CE) )^2+ {\rm Ch}_{\boldsymbol{Y}}\big(\CE^2\big) \big), \\
{\rm Ch}_{\tableau{1 1},\boldsymbol{Y}}(\CE) ={1\over 2} \big( ( {\rm Ch}_{\boldsymbol{Y}}(\CE) )^2- {\rm Ch}_{\boldsymbol{Y}}\big(\CE^2\big) \big).
\end{gather*}
We then define
\begin{gather*}
W_\CR ={1\over Z_N} \sum_{\boldsymbol{Y}} \big((-1)^N\Lambda_{\rm 5d}^{2N} \big)^{\ell (\boldsymbol{Y})} {\rm Ch}_{\CR,\boldsymbol{Y}}(\CE) z_{\boldsymbol{Y}},
\end{gather*}
which are indeed of the form
\begin{gather*}
W_\CR=s_\CR\big( \re^{\alpha_I}\big)+\cdots,
\end{gather*}
as expected. The relation (\ref{qmm}) is then given by the Wilson loops in purely antisymmetric representations,{\samepage
\begin{gather*}
H_k= W_{(1^k)}, \qquad k=1, \dots, N-1,
\end{gather*}
where $(1^k)$ is the vertical Young tableau with $k$ boxes.}

In order to proceed, it is useful to introduce some ingredients from the Lie algebra of ${\rm SU}(N)$. We will write the moduli as elements of the vector space $\Lambda_{\rm w} \otimes \IC$, where $\Lambda_{\rm w}$ is the weight lattice of ${\rm SU}(N)$. We recall that this vector space has a scalar product induced by the Cartan--Killing form. The fundamental weights $\{ \blam_i\}_{i=1, \dots, N-1}$ are a basis of $\Lambda_{\rm w}$. A useful set of generators is given by the weights of the fundamental representation, which we will denote by $\bs{e}_j$, $j=1, \dots, N$. These weights can be written, in terms of the fundamental weights, as
\begin{gather*}
\bs{e}_1 = \blam_1, \qquad \bs{e}_i = -\blam_{i-1}+ \blam_i, \qquad 2 \le i\le N-1,\qquad \bs{e}_N = -\blam_{N-1},
\end{gather*}
 and they satisfy
\begin{gather*}
 \bs{e}_i \cdot \bs{e_j}= \delta_{ij}-{1\over N}, \qquad i,j=1, \dots, N.
\end{gather*}
 The simple roots are given by\footnote{These should not be confused with the parameters introduced in~(\ref{Qal}). Simple roots are always written in boldface letters.}
\begin{gather*}
\balpha_i =\bse_i-\bse_{i+1}, \qquad i=1, \dots, N-1.
\end{gather*}
We now write the vector of moduli as
\begin{gather*}
\bs{a}=\sum_{j=1}^N \alpha_j \bs{e}_j.
\end{gather*}
We can also introduce a useful parametrization of the moduli in terms of $N-1$ independent quantities, $a_j$, $j=1, \dots, N-1$, as follows
\begin{gather*}
\bs{a}= \sum_{j=1}^{N-1} a_j \bs{\lambda}_j,
\end{gather*}
so that
\begin{gather*}
a_i=\alpha_i-\alpha_{i+1}, \qquad i=1, \dots, N-1.
\end{gather*}
In addition, we introduce a vector of derivatives as
\begin{gather*}
{\partial \over \partial \bs{a}}= \sum_{j=1}^{N-1} \bs{\alpha}_j {\partial \over \partial a_j } = \sum_{j=1}^N \bs{e}_j {\partial \over \partial \alpha_j } .
\end{gather*}
We will denote the Weyl group of the Lie algebra $A_{N-1}$ by $\CW_N$. $\CW_N$ is isomorphic to the permutation group $S_N$, and it acts on the weight lattice $\Lambda_{\rm w}$ by permutation of the labels in the weights $\bs{e}_i$, $i=1, \dots, N$. In particular, it induces a permutation of the indices in the moduli $\alpha_I$.

Let us now consider the 4d limit of the 5d quantities introduced above. To do that, we introduce an explicit five-dimensional radius $R$ as
\begin{gather*}
\epsilon_j \rightarrow R \epsilon_j, \qquad \alpha_I \rightarrow R \alpha_I, \qquad \Lambda_{\rm 5d}= R \Lambda,
\end{gather*}
and we consider the limit
\begin{gather}\label{R4d}
R \rightarrow 0.
\end{gather}
The spectral curve (\ref{rt-sc}) becomes, after scaling $x\rightarrow R x$ and $\zeta_I \rightarrow R\zeta_I$,
\begin{gather}\label{rt-sc-N}
R^N\Lambda^N \big( \re^p + \re^{-p} \big)+ \prod_{i=1}^{N} 2 \sinh \left( R {x- \zeta_I \over 2}\right)=0,
\end{gather}
and in the limit (\ref{R4d}) we recover (\ref{curve}). The partition function in 4d is then given by
\begin{gather}\label{z4d}
Z (\bs{a}; \epsilon_1, \epsilon_2)=\sum_{\boldsymbol{Y}} \big((-1)^N \Lambda^{2N} \big)^{\ell (\boldsymbol{Y})} \CZ_{\boldsymbol{Y}},
\end{gather}
where
\begin{gather*}
\CZ_{\bs{Y}} =\prod_{I,J=1}^N \prod_{s \in Y_I} {1\over \alpha_I -\alpha_J -\epsilon_1 v_{Y_J}(s) +\epsilon_2 ( h_{Y_I}(s)+1 )} \\
\hphantom{\CZ_{\bs{Y}} =}{} \times \prod_{s\in Y_J} {1\over \alpha_I -\alpha_J +\epsilon_1 (v_{Y_I}(s)+1){-} \epsilon_2 h_{Y_J}(s)}.
\end{gather*}
The NS free energy in 4d is defined as
\begin{gather}\label{fns4d}
F_{\rm NS}^{\rm inst}(\bs{a}, \hbar) =\ri \hbar \lim_{\epsilon_2 \rightarrow 0} \epsilon_2 \log Z(\bs{a}; \ri \hbar , \epsilon_2).
\end{gather}
For example, for ${\rm SU}(3)$ one finds
\begin{gather*}
F_{\rm NS}^{\rm inst}(a_1, a_2, \hbar)=-\frac{2 \Lambda^6 \big(a_1^2+a_1 a_2+a_2^2+3 \hbar ^2\big)}{\big(a_1^2+\hbar ^2\big)
\big(a_2^2+\hbar ^2\big) \big((a_1+a_2)^2+\hbar ^2\big)}+\CO\big(\Lambda^{12} \big).
\end{gather*}
As our notation indicates, (\ref{fns4d}) is just the instanton part of the NS free energy. The total NS free energy involves as well a perturbative part. This part can be unambiguously obtained from the 5d perspective, as shown in~\cite{hm}. The building block of this part is the function,
\begin{gather}\label{gam4d}
 \gamma_{\rm 4d}(a, \hbar ) = a \log \left( { \hbar \over \Lambda} \right)-{\pi \hbar \over 4} -{\ri \hbar \over 2} \log {\Gamma ( 1+\ri a /\hbar ) \over \Gamma ( 1-\ri a /\hbar)}.
\end{gather}
A closely related function appears in \cite{no2}. We will denote by $\Delta= \Delta_+\cup \Delta_-$ the set of roots of the Lie algebra, where $\Delta_{\pm}$ is the set of positive (respectively, negative) roots. We define the total NS free energy (including perturbative and instanton corrections) through the equation
\begin{gather*}
{\partial F_{\rm NS} \over \partial \bs{a}}= {\partial F^{\rm inst}_{\rm NS} \over \partial \bs{a}}+ 2 \sum_{\balpha \in \Delta_+} \gamma_{\rm 4d}\left( \bs{a}\cdot \balpha, \hbar \right)\balpha.
\end{gather*}
We note that
\begin{gather*}
\sum_{\balpha \in \Delta_-} \gamma_{\rm 4d} ( \bs{a}\cdot \balpha, \hbar )\balpha=\pi \hbar \bs{\rho}+ \sum_{\balpha \in \Delta_+} \gamma_{\rm 4d}( \bs{a}\cdot \balpha, \hbar)\balpha,
\end{gather*}
 where
\begin{gather*}
\bs{\rho}= {1\over 2} \sum_{\balpha \in \Delta_+} \balpha=\sum_{i=1}^{N-1} \blam_i =\sum_{i=1}^N (N-i) \bs{e}_i
\end{gather*}
is the Weyl vector, so that we can also write
\begin{gather*}
 {\partial F_{\rm NS} \over \partial \bs{a}}= {\partial F^{\rm inst}_{\rm NS} \over \partial \bs{a}}+\sum_{\balpha \in \Delta} \gamma_{\rm 4d} ( \bs{a}\cdot \balpha, \hbar )\balpha -\pi \hbar \bs{\rho}.
\end{gather*}

The relation (\ref{qmm}) in 5d induces a corresponding one in 4d, of the form
\begin{gather}\label{4drels}
h_k= h_k(\bs{a}; \epsilon_1, \epsilon_2), \qquad k=2, \dots, N.
\end{gather}
In the NS limit, these relations become the 4d quantum mirror map, which we will write as
\begin{gather}\label{4dqmm}
h_k=h_k(\bs{a}, \hbar).
\end{gather}
This is a resummed or exact version of the inverse function to the quantum $A$-periods appearing in the first line of (\ref{qpers-int}). The explicit form of the relations (\ref{4drels}) can be obtained by expanding the 5d functions~(\ref{qmm}) in powers of~$R$ around $R=0$. Explicitly, this goes as follows. Let us introduce
\begin{gather*}
c_2(Y)= {1\over 2} \sum_{i\ge1}y_i (y_i-1).
\end{gather*}
Then, the character (\ref{cher-char}) has the expansion,
\begin{gather}\label{ch-r}
{\rm Ch}_{\boldsymbol{Y}}(\CE)= N +\sum_{k=0}^\infty \CC_k(\bs{a}, \bs{Y}) R^k ,
\end{gather}
where the very first $\CC_k$ are
\begin{gather}\label{ccs}
\CC_2(\bs{a}, \bs{Y}) ={1\over 2} \sum_{I=1}^N \alpha_I^2 -\epsilon_1 \epsilon_2 \ell (\boldsymbol{Y}) , \\
\CC_3(\bs{a}, \bs{Y}) = \epsilon_1 \epsilon_2 \left( {\epsilon_1 +\epsilon_2 \over 2} \ell (\boldsymbol{Y}) +\epsilon_1\sum_{I=1}^N c_2(Y^t_I) +\epsilon_2\sum_{I=1}^N c_2(Y_I) - \sum_{I=1}^N \alpha_I \ell(Y_I) \right)
+ {1\over 6} \sum_{I=1}^N \alpha_I^3.\nonumber
\end{gather}
There are similar expansions for the characters in higher representations, which can be derived from (\ref{ch-r}). Through (\ref{wl-fund}) one derives an expansion for $H_1$ in powers of $R$. On the other hand, this expansion has the form
\begin{gather*}
 H_1= N +\sum_{j\ge 2} {R^j \over j!} p_j,
\end{gather*}
where the $p_j$ are the quantum analogues of Newton polynomials, and they can be related to the $h_k$ by using standard results in the theory of symmetric functions. The general Newton polynomial $p_{\bs{k}}$ is labelled by a vector $\bs{k}=(k_1, k_2,\dots)$, and is given by
\begin{gather*}
 p_{\bs{k}}= \prod_{j\ge 1} p_j^{k_j}.
\end{gather*}
 Let us denote
\begin{gather*}
 |\bs{k}|= \sum_{j\ge 1} k_j.
\end{gather*}
 Then \cite{macdonald},
\begin{gather*}
h_n= s_{(1^n)}= \sum_{\bs{k}} {\epsilon(\bs{k}) \over z_{\bs{k}}} p_{\bs{k}},
\end{gather*}
 where $z_{\bs{k}}$ has been defined in (\ref{zk}), and
\begin{gather*}
 \epsilon( \bs{k})= (-1)^{ |\bs{k}|- \ell(\bs{k})}.
\end{gather*}
 Note that $p_1=0$, so that only vectors of the form $(0, k_2,\dots)$ contribute in the above sum. We find, for example,
\begin{gather*}
 h_2 =-{1\over 2} p_2, \qquad h_3 = {1\over 3} p_3, \qquad h_4 = {1 \over 8}p_2^2- {1\over 4} p_4.
\end{gather*}
One finds in this way, for example (see also \cite{fm-wilson}),
\begin{gather}
h_2(\bs{a}; \epsilon_1, \epsilon_2 t)= -{1\over Z} \sum_{\bs{Y}} \big((-1)^N \Lambda^{2N} \big)^{\ell (\boldsymbol{Y})} \CC_2 (\bs{a}, \bs{Y}) \CZ_{\boldsymbol{Y}},\nonumber\\
h_3 (\bs{a}; \epsilon_1, \epsilon_2 ) = {2\over Z} \sum_{\bs{Y}} \big((-1)^N \Lambda^{2N} \big)^{\ell (\boldsymbol{Y})} \CC_3 (\bs{a}, \bs{Y}) \CZ_{\boldsymbol{Y}}.\label{qmm-ex}
\end{gather}
It follows from the first line in (\ref{ccs}) that $h=-h_2$ is given by
\begin{gather}\label{mat1}
h (\bs{a}; \epsilon_1, \epsilon_2 )={1\over 2}\sum_{I=1}^N \alpha_I^2 -\epsilon_1 \epsilon_2 \Lambda^{2N} {\partial \over \partial \Lambda^{2N}} \log Z (\bs{a}; \epsilon_1, \epsilon_2 ).
\end{gather}
In particular, in the NS limit we have
\begin{gather}\label{mat2}
h (\bs{a}, \hbar )={1\over 2}\sum_{I=1}^N \alpha_I^2 - \Lambda^{2N} {\partial \over \partial \Lambda^{2N}} F_{\rm NS}^{\rm inst}(\bs{a}, \hbar).
\end{gather}
The relations (\ref{mat1}), (\ref{mat2}) generalize Matone's relation \cite{matone} to the Omega-background~\cite{francisco}.

An important observation is that instanton calculus gives the NS free energy and the quantum mirror maps as infinite sums in the instanton counting parameter $\Lambda^{2N}$. When $\hbar=0$, these infinite sums are convergent in the so-called large radius or semiclassical region, in which the $a_i$'s are large. As it is well-known in SW theory (and similar to what happens in mirror symmetry), there is a natural boundary for the region of convergence, which is the discriminant locus of the SW curve. Near this locus, convergence breaks down, and one needs to perform an analytic continuation to have access to other regions in moduli space. This should remain qualitatively true when $\hbar \not=0$, namely, the expansions (\ref{z4d}) and (\ref{qmm-ex}) should have a finite radius of convergence around the region at infinity in moduli space. Although the precise analytic structure of the resulting functions is not known, numerical calculations seem to validate this picture (some analyticity results for the instanton partition function have been derived in \cite{felder}, but they do not apply to the case $\hbar >0$ in which we are interested). This expected analyticity is of paramount importance, since it means that the quantum periods, defined as instanton sums in gauge theory, are well-defined functions,
at least in some region of moduli space, and Borel-type resummations are no longer necessary.

It was pointed out in \cite{ns}, and further elaborated in \cite{bourgine, kt2,my}, that the NS free energy and the quantum mirror maps can be also computed in terms of a set of integral equations reminiscent of TBA equations. In this approach, one first introduces a function $\varphi(x; \bs{\alpha})$ as a~solution of the following integral equation,
\begin{gather*}
\varphi(x;\bs{a})=-\int_{\IR} \frac{\rd y}{2\pi} K(x-y) \log\big(1+\Lambda^{2N} Q(y;\bs{a}) \re^{-\varphi(y;\bs{a})} \big),
\end{gather*}
where
\begin{gather*}
K(x) =\frac{2\hbar}{x^2+\hbar^2},\qquad
Q(x;\bs{a}) =\prod_{j=1}^N \frac{1}{(x-\alpha_j-\ri \hbar/2)(x-\alpha_j+\ri \hbar/2)}.
\end{gather*}
The NS free energy is then given by
\begin{gather*}
F_\text{NS}^\text{inst}(\bs{a},\hbar) =-\frac{\hbar}{2\pi} \int_{\IR} \rd x \biggl[ {-}\frac{1}{2} \varphi(x;\bs{a}) \log\big(1+\Lambda^{2N} Q(x;\bs{a}) \re^{-\varphi(x;\bs{a})}\big)\\
\hphantom{F_\text{NS}^\text{inst}(\bs{a},\hbar) =-\frac{\hbar}{2\pi} \int_{\IR} \rd x}{} +{\rm Li}_2\big({-}\Lambda^{2N} Q(x;\bs{a}) \re^{-\varphi(x;\bs{a})} \big) \biggr].
\end{gather*}
From these expressions one can also compute the partial derivatives of the NS free energy as
\begin{gather*}
\frac{\partial F_\text{NS}^\text{inst}}{\partial \bs{a}}
=\frac{\hbar}{2\pi} \int_{\IR} \rd x {\partial \log Q (x; \bs{a}) \over \partial \bs{a}}\log\big(1+\Lambda^{2N} Q(x;\bs{a}) \re^{-\varphi(x;\bs{a})} \big). 
\end{gather*}
One can also use the TBA integral equations to compute the quantum mirror map. The result is more easily formulated in terms of the Newton polynomials $p_k$:
\begin{gather*}
p_k = \sum_{I=1}^N \alpha_I^k +{k \over 2 \pi \ri} \int_\IR \rd x \big[ (x+ \ri \hbar/2)^{k-1}- (x- \ri \hbar/2)^{k-1}\big] \log\big(1+\Lambda^{2N} Q(x;\bs{a}) \re^{-\varphi(x;\bs{a})} \big),
\end{gather*}
for $k=2, \dots, N$. One finds in this way, for example,
\begin{gather*}
h ={1\over 2} \sum_{I=1}^N \alpha_I^2+\frac{\hbar}{2\pi} \int_{-\infty}^\infty \rd x \log\big(1+\Lambda^{2N} Q(x;\bs{a}) \re^{-\varphi(x;\bs{a})} \big), \\
h_3 ={1\over 3} \sum_{I=1}^N \alpha_I^3+\frac{\hbar}{\pi} \int_{-\infty}^\infty \rd x x \log\big(1+\Lambda^{2N} Q(x;\bs{a}) \re^{-\varphi(x;\bs{a})} \big).
\end{gather*}
It is natural to ask whether and where the TBA equations provide well-defined functions for the NS free energy and the quantum mirror maps, complementing in this way the approach based on instanton calculus in gauge theory. The results of~\cite{kt2} guarantee that these quantities are well defined in a region of moduli space containing the Toda lattice points. Numerical studies support this conclusion, but it would be of course important to know if this region can be enlarged, maybe to the whole of moduli space.

We conclude this summary by emphasizing that supersymmetric gauge theory, in the NS limit, provides a very powerful resummation of the traditional quantum periods appearing in the all-orders WKB method, as applied to the SW curve. This will allow us to define exact quantization conditions for the spectral problem (\ref{specp2}) involving actual functions, and not resurgent functions.

\section{The exact quantization condition}\label{sec:eqc}

\subsection{Statement and properties}

We are now ready to state the exact quantization condition for the spectral problem (\ref{specp1}) (or, equivalently, (\ref{specp2})). Let us consider the following vector in the weight lattice $\Lambda_w$,
\begin{gather}\label{spec-vector}
\bs{\gamma}= \sum_{i=1}^{N-1} (-1)^{i-1} \bs{\lambda}_i={1\over 2}\sum_{i=1}^N (-1)^{i-1} \bs{e}_i.
\end{gather}
The Weyl orbit of $\bs{\gamma}$ is defined by
\begin{gather}\label{orbit}
\CW_N\cdot \bs{\gamma} =\big\{ w(\bs{\gamma})\colon w\in \CW_{N} \big\},
\end{gather}
i.e., it consists of all the vectors in $\Lambda_{\rm w}$ obtained by acting with an element of the Weyl group $\CW_N$ on $\bs{\gamma}$. Since the Weyl group acts as the permutation group $S_N$ on the indices of the weights~$\bs{e}_i$, the number of different elements in the Weyl orbit will be
\begin{gather*}
d^{\rm e}_N = {N \choose N/2}
\end{gather*}
when $N$ is even, and
\begin{gather*}
d^{\rm o}_N= {N \choose {N-1 \over 2}}
\end{gather*}
when $N$ is odd.

There are two different cases for the exact quantization conditions, depending on the parity of $N$. When $N$ is even, the quantization condition determining the spectrum of bound states is
\begin{gather}\label{eqc-even}
\boxed{\sum_{\bs{n} \in \CW_N\cdot \bs{\gamma} } \exp\left( {\ri \over \hbar} {\partial F_{\rm NS} \over \partial \bs{a}}\cdot \bs{n} \right) \prod_{\alpha \in \Delta_+}
 \left( 2 \sinh \left( {\pi \bs{a}\cdot \balpha\over \hbar}\right) \right)^{-\left(\bs{n}\cdot \balpha\right)^2} =0.}
\end{gather}
When $N$ is odd, the quantization condition determining the spectrum of resonances is
\begin{gather}\label{eqc-odd}
\boxed{\sum_{\bs{n} \in \CW_N\cdot \bs{\gamma} } \exp\left( {\ri \over \hbar} {\partial F_{\rm NS} \over \partial \bs{a}}\cdot \bs{n} -{\pi \over \hbar}\bs{a} \cdot \bs{n}\right) \prod_{\alpha \in \Delta_+} \left( 2 \sinh \left( {\pi \bs{a}\cdot \balpha\over \hbar}\right) \right)^{-\left(\bs{n}\cdot \balpha\right)^2} =0.}
\end{gather}
As we mentioned above, these quantization conditions provide one single constraint on the values of the moduli. To solve the spectral problem~(\ref{specp2}), we fix the $N-2$ coefficients $h_2, \dots, h_{N-1}$ in the
potential $V_N(x)$. The quantization condition and the quantum mirror map~(\ref{4dqmm}) for these $N-2$ coefficients give in total $N-1$ conditions for~$N-1$ independent variables~$a_i$, $i=1, \dots, N-1$. The values of these variables, once plugged into the quantum mirror map for~$h_N$, finally give the eigenvalue for the spectral problem.

The quantization conditions (\ref{eqc-even}), (\ref{eqc-odd}) have two important formal properties: they are invariant under the action of the Weyl group on $\bs{a}$, and they are solved by the points in the moduli space corresponding to the spectrum of the periodic Toda lattice, as expected from our discussion in Section~\ref{sec:sp}. Let us first discuss invariance under the Weyl group. This is only expected, since two points $\bs{a}$, $\bs{a}'=w(\bs{a})$ related by an element of the Weyl group represent the same point in moduli space. To check the invariance, let us suppose that we act with an element $w \in \CW_{N}$ on $\bs{a}$. We have
\begin{gather*}
\sum_{\balpha \in \Delta} \gamma_{\rm 4d} ( w(\bs{a})\cdot \balpha, \hbar )\balpha =
 \sum_{\balpha \in \Delta} \gamma_{\rm 4d} \big( \bs{a}\cdot w^{-1} ( \balpha ), \hbar \big)\balpha = \sum_{\balpha \in \Delta} \gamma_{\rm 4d} ( \bs{a}\cdot \balpha , \hbar )w(\balpha),
\end{gather*}
where we used invariance of $\Delta$ and of the Cartan--Killing form under Weyl permutations. We conclude that, under $w \in \CW_{N}$, one has
\begin{gather*}
{\partial F_{\rm NS} \over \partial \bs{a}} \rightarrow w\left( {\partial F_{\rm NS} \over \partial \bs{a}} + \pi \hbar \bs{\rho} \right)- \pi \hbar \bs{\rho}.
\end{gather*}
We recall that, under the action of an element $w$ of the Weyl group, the positive roots are transformed into positive or negative roots (for this and other statements on root and weight lattices, see, e.g., \cite{humphreys} or \cite[Chapter~13]{yellowbook}). Therefore, we can write
\begin{gather*}
 w(\balpha)= \epsilon(w, \balpha) \balpha', \qquad \balpha'\in \Delta_+, \qquad \epsilon(w, \balpha) =\pm 1,
\end{gather*}
and
\begin{gather*}
 \prod_{\alpha \in \Delta_+} \left( 2 \sinh \left( {\pi w(\bs{a})\cdot \balpha\over \hbar}\right) \right)^{-(\bs{n}\cdot \balpha)^2} =
 \prod_{\alpha \in \Delta_+} \left( 2 \sinh \left( {\pi \bs{a}\cdot w^{-1} \left( \balpha \right) \over \hbar}\right) \right)^{-(\bs{n}\cdot \balpha )^2} \\
\hphantom{\prod_{\alpha \in \Delta_+} \left( 2 \sinh \left( {\pi w(\bs{a})\cdot \balpha\over \hbar}\right) \right)^{-(\bs{n}\cdot \balpha)^2}}{}
 = \epsilon\big(w^{-1}, \bs{n}\big) \prod_{\alpha \in \Delta_+} \left( 2 \sinh \left( {\pi \bs{a}\cdot \balpha\over \hbar}\right) \right)^{-\left(w^{-1}(\bs{n})\cdot \balpha\right)^2},
\end{gather*}
where
\begin{gather*}
\epsilon(w, \bs{n})= \prod_{\alpha \in \Delta_+} ( \epsilon(w, \balpha))^{(\bs{n}\cdot \balpha)^2}
\end{gather*}
is a sign depending on $w$ and $\bs{n}$. The l.h.s.\ of the quantization condition becomes, for $N$ even,
\begin{gather*}
\sum_{\bs{n} \in \CW_N\cdot \bs{\gamma}} \epsilon\big(w^{-1}, \bs{n}\big) \re^{\pi \ri \bs{\rho} \cdot \left(w^{-1}(\bn) - \bn\right)} \exp\left( {\ri \over \hbar} {\partial F_{\rm NS} \over \partial \bs{a}}\cdot w^{-1}(\bs{n}) \right)\\
\qquad {}\times \prod_{\alpha \in \Delta_+} \left( 2 \sinh \left( {\pi \bs{a}\cdot \balpha\over \hbar}\right) \right)^{-\left(w^{-1}(\bs{n})\cdot \balpha\right)^2}.
\end{gather*}
Since we are summing over a Weyl orbit, this is invariant provided
\begin{gather}\label{cond-inv}
 \epsilon(w, \bs{n}) \re^{\pi \ri \bs{\rho} \cdot (w(\bn) - \bn)} =1,
\end{gather}
 for all $w\in \CW_N$ and all $\bn$ appearing in the sum. This can be verified by considering the generators of the Weyl group $w_i$ $i=1, \dots, N-1$, which are reflections associated to simple roots. These reflections leave invariant all the positive roots, except~$\balpha_i$, which changes sign. Then,
\begin{gather*}
 \epsilon(w_i, \bs{n}) = (-1)^{\bn \cdot \balpha_i} ,
\end{gather*}
 where we have used the fact, easily checked, that, for any $\balpha \in \Delta$, $\bn \cdot \balpha$ can only take the va\-lues~$\pm1$,~$0$. At the same time
\begin{gather*}
 \bs{\rho} \cdot (w_i(\bn) - \bn)=-(\bn \cdot \balpha_i) \bs{\rho}\cdot \balpha_i=-\bn \cdot \balpha_i,
\end{gather*}
 so that
\begin{gather*}
 \re^{\pi \ri \bs{\rho} \cdot (w(\bn) - \bn)}=(-1)^{\bn \cdot \balpha_i},
\end{gather*}
and (\ref{cond-inv}) holds. A similar argument holds for $N$ odd.

It can be explicitly checked that the points $\bs{a}$ in the spectrum of the Toda lattice sa\-tis\-fy~(\ref{eqc-even}),~(\ref{eqc-odd}). These points are characterized by the quantization conditions \cite{kt2, ns}
\begin{gather}\label{toda-qc}
{1 \over \hbar} {\partial F_{\rm NS} \over \partial \bs{a}}= 2 \pi \bs{\ell}+ \pi \bs{\rho},
\end{gather}
where
\begin{gather*}
 \bs{\ell}= \sum_{k=1}^{N-1} \ell_k \blam_k, \qquad \ell_k \in \IZ_{\ge 0}.
\end{gather*}
To verify this property, we have to evaluate
\begin{gather*}
{\ri \over \hbar} {\partial F_{\rm NS} \over \partial \bs{a}}\cdot (w(\bs{\gamma})- \bs{\gamma})= 2 \pi \ri \bs{\ell} \cdot (w(\bs{\gamma})- \bs{\gamma}) + \pi \ri \bs{\rho} \cdot (w(\bs{\gamma})- \bs{\gamma}).
\end{gather*}
The first term in the r.h.s.\ belongs to $2 \pi \ri \IZ$. This is because, since $\bs{\gamma}$ is a weight,
\begin{gather*}
w(\bs{\gamma})- \bs{\gamma} \in \Lambda_{\rm r}
\end{gather*}
and its product with an element in $\Lambda_{\rm w}$ like $\bs{\ell}$ is an integer. The condition that points in the spectrum of the Toda lattice satisfy the quantization condition is
\begin{gather*}
\sum_{w\in \CW_{N}} \re^{\pi \ri \bs{\rho}\cdot w(\bs{\gamma})} \prod_{\alpha \in \Delta_+} \left( 2 \sinh \left( {\pi \bs{a}\cdot \balpha\over \hbar}\right) \right)^{-\left(w(\bs{\gamma})\cdot \balpha\right)^2} =0,
\end{gather*}
for $N$ even, and
\begin{gather*}
 \sum_{w\in \CW_{N}} \re^{(\pi \ri \bs{\rho} -\pi \bs{a} /\hbar)\cdot w(\bs{\gamma})} \prod_{\alpha \in \Delta_+} \left( 2 \sinh \left( {\pi \bs{a}\cdot \balpha\over \hbar}\right) \right)^{-\left(w(\bs{\gamma})\cdot \balpha\right)^2} =0,
\end{gather*}
for $N$ odd. Although we do not have a proof of this fact, we have verified up to high values of~$N$ that the above equalities are satisfied by any~$\bs{a}$.

Our quantization conditions (\ref{eqc-even}), (\ref{eqc-odd}) are very similar to what is obtained with the exact WKB method in ordinary quantum mechanics. In this case, the relevant curve is (in appropriate units)
\begin{gather}\label{wkb-curve}
p^2=E-V(x),
\end{gather}
where $E$ is the energy, $p$ is the momentum, and $V(x)$ is the potential. The all-orders WKB method applied to (\ref{wkb-curve}) makes it possible to define quantum periods, similar to (\ref{qpers-int}). These periods are generically complex and they inherit the integral lattice structure of the space of one-cycles on the curve. One obtains in this way an integer lattice $\Lambda_{\rm WKB}$ of rank~$2g$, where~$g$ is the genus of the curve (\ref{wkb-curve}). In the seminal work of Balian, Parisi and Voros~\cite{bpv} it was argued that in principle all the periods in the lattice $\Lambda_{\rm WKB}$ play a r\^ole in the quantum theory. In particular, the exact quantization condition will involve complex periods which are invisible in the classical theory. This was illustrated in~\cite{bpv} and further investigated in \cite{voros,voros-quartic} in the example of the pure quartic oscillator with Hamiltonian $\mH= \mm^2 + \mx^4$. In the classical theory, the motion takes place in the interval $\big[{-}E^{1/4}, E^{1/4}\big]$. The {\it perturbative} WKB condition involves the quantum period obtained by integrating around this interval. However, the quantum period obtained by integration around the interval $\big[{-}\ri E^{1/4}, \ri E^{1/4}\big]$, although it is classically invisible, gives a {\it non-perturbative correction} which is crucial to reproduce the correct spectrum.

More generally, quantization conditions obtained with the exact WKB method in ordinary quantum mechanics involve a sum of terms which is then set to zero. Each term in this sum is given by a numerical coefficient, times a {\it Voros multiplier} \cite{ddpham,dpham,reshyper}. We recall that a Voros multiplier has the form
\begin{gather*}
\re^{\ri \CA/\hbar}, \qquad \CA \in \Lambda_{\rm WKB}.
\end{gather*}
For example, the quantization condition for the resonances of the standard cubic oscillator can be written as \cite{alvarez-casares2, ddpham}:
\begin{gather}\label{cubic-q}
1+ \re^{ \ri \nu / \hbar}+ \re^{-{\partial F \over \partial \nu}/\hbar}=0,
\end{gather}
where we have used the notation in \cite{cm-ha} to emphasize the similarities with the case studied in this paper. In this notation, $\nu$ is the quantum period around the perturbative cycle of the cubic oscillator, while $\ri \partial F/\partial \nu$ is the quantum period around the tunneling cycle.

The quantization conditions (\ref{eqc-even}), (\ref{eqc-odd}) can be also written as a sum of Voros multipliers. These multipliers involve integer linear combinations of the quantum periods of the SW curve:
\begin{gather*}
{\partial F_{\rm NS} \over \partial a_j}, \qquad \pi \ri a_j, \qquad j=1, \dots, N-1 .
\end{gather*}
 The quantum $B$-periods are the ones that appear in the quantization condition of the Toda lattice. They correspond to the cycles around the intervals of instability in the spectral curve. The contribution from the quantum $A$-periods is new, and as we will see in Section~\ref{derivation}, their origin is purely non-perturbative: they are the remnants in 4d of the non-perturbative correction due to the standard topological string in the TS/ST correspondence. In contrast to the solution of the spectral problem of the Toda lattice, which involves only the ``perturbative'' $B$-periods, the solution of the spectral problem associated to the SW curve requires these new, non-perturbative Voros multipliers.

Although our quantization conditions are very similar to the ones appearing in the exact WKB method, there is as well an important difference: in quantization conditions like (\ref{cubic-q}), the quantum periods are {\it formal} power series, which have to be resummed with Borel--\'Ecalle techniques in order to obtain the actual spectra. In particular, the form of the quantization conditions can change as we cross Stokes lines (this is in fact the case for the cubic oscillator, and the form (\ref{cubic-q}) is only valid for a specific lateral resummation, see \cite{alvarez-casares2} for more details). In contrast, our quantization conditions (\ref{eqc-even}), (\ref{eqc-odd}) involve resummed versions of these periods, which can be defined as convergent instanton sums in gauge theory or as well-defined solutions to TBA equations. Therefore, our quantization conditions are truly exact, in the sense that they involve actual functions.

As we already anticipated, the quantization conditions \eqref{eqc-even}, \eqref{eqc-odd} can be derived from the TS/ST duality \cite{cgm,ghm} which was originally formulated for real values of $\hbar$. However it has been shown in \cite{gm17} that this correspondence can be extended to complex values as well. We have tested that \eqref{eqc-even}, \eqref{eqc-odd} also holds for complex values of $\hbar$ as far as ${\rm Re}(\hbar) \neq 0$.

\subsection{Examples and evidence}

\subsubsection[The case $N=2$]{The case $\boldsymbol{N=2}$}
Let us first consider the case $N=2$. The vector (\ref{spec-vector}) is $\bs{\gamma}= \bs{\lambda}_1$, and the Weyl orbit is
\begin{gather*}
\CW_2 \cdot \bs{\gamma}= \{ \bs{\lambda}_1, - \bs{\lambda}_1 \},
\end{gather*}
so it contains two elements. Setting $a_1=a$, we find the quantization condition
\begin{gather*}
\cos \left( {1\over \hbar} {\partial F_{\rm NS} \over \partial a} \right)=0.
\end{gather*}
This is the quantization condition found in \cite{ns} for $N=2$ Toda lattice. In this case the integrable system has a single Hamiltonian, which equivalent to our~$\mH_2$. Therefore our exact quantization condition reproduces the expected result.

\subsubsection[The case $N=3$]{The case $\boldsymbol{N=3}$}
Let us then consider the next case, $N=3$. The Hamiltonian is given in (\ref{n3hamil}). The vector (\ref{spec-vector}) is
\begin{gather*}
\bs{\gamma}= {1\over2} ( \bs{e}_1 -\bs{e}_2+\bs{e}_3 ),
\end{gather*}
and its Weyl orbit contains three elements:
\begin{gather*}
\CW_3\cdot \bs{\gamma}= \left\{{1\over2} ( \bs{e}_1 -\bs{e}_2+\bs{e}_3 ), {1\over2} ( -\bs{e}_1 +\bs{e}_2+\bs{e}_3 ), {1\over2} ( \bs{e}_1 +\bs{e}_2-\bs{e}_3 ) \right\}.
\end{gather*}
Let us introduce the functions $\phi_{1,2}(a_1, a_2, \hbar)$, defined as
\begin{gather*}
 \phi_1(a_1, a_2; \hbar)={1\over \hbar} \left( {\partial F_{\rm NS} \over \partial a_2} - 2 {\partial F_{\rm NS} \over \partial a_1} \right), \qquad
 \phi_2(a_1, a_2; \hbar)= {1\over \hbar} \left(2 {\partial F_{\rm NS} \over \partial a_2} - {\partial F_{\rm NS} \over \partial a_1} \right).
\end{gather*}
Then, elementary algebra shows that the quantization condition can be written as
\begin{gather}\label{n3eqc}
1+ {1 -\re^{-2 \pi a_1/\hbar} \over 1- \re^{-2 \pi (a_1+a_2) /\hbar}}\re^{-2 \pi a_2/\hbar} \re^{\ri \phi_2} +{1 -\re^{-2 \pi a_2/\hbar} \over 1- \re^{-2 \pi (a_1+a_2) /\hbar}} \re^{\ri \phi_1}=0.
\end{gather}
It is easy to check directly that this condition is satisfied by the points belonging to the spectrum of the Toda lattice, since the quantization conditions (\ref{toda-qc}) imply that
\begin{gather*}
\re^{\ri \phi_j}=-1, \qquad j=1,2.
\end{gather*}
When this holds, the l.h.s.\ of (\ref{n3eqc}) vanishes identically.

\begin{table}[tb]\centering
 \begin{tabular}{cc}\hline
	Instantons & $h^{(0)}_3$ \\ \hline
	 $0$ & $-\underline{0.06}40367918244+\underline{1.0}825753128 \cdot 10^{-8} \ri $ \\
	 $3$ & $-\underline{0.0637695}052176+\underline{1.07506}20426 \cdot 10^{-8} \ri$ \\
	 $5$ & $-\underline{0.0637695528}598+\underline{1.075063257}5\cdot 10^{-8} \ri$ \\ \hline
	 TBA & $-0.0637695528692+1.0750632578\cdot 10^{-8} \ri$\\ \hline
	 numerical & $-0.0637695528691904+1.0750632578 \cdot 10^{-8} \ri$ \\ \hline
 \end{tabular}
 \caption{Ground state of $\mH_3$ for $h=5/2 (15/2)^{1/3}$. We set $\hbar=\Lambda=1$. The number appearing in the column labelled as ``Instantons'' refers to the order at which we truncate the series~\eqref{z4d} and~\eqref{qmm-ex}.}\label{tab:spectraN3-1}
\end{table}

Let us now test (\ref{n3eqc}). To do this, we compute numerically the spectrum of $\mH_3$ by using the complex dilatation method combined with the standard Rayleigh--Ritz method. In practice, we perform a numerical diagonalization of the rotated Hamiltonian~(\ref{Htheta}) in the basis of eigenstates of the harmonic oscillator. This gives the spectrum of complex resonances for $h_3$, $h_3^{(n)}$, $n=0,1,2,\dots$, as a function of $h=-h_2$. On the other hand, we solve simultaneously~(\ref{n3eqc}) and the equation
\begin{gather*}
h=- h_2(a_1, a_2; \hbar)
\end{gather*}
for the chosen value of $h$. This produces a discrete sequence of values for $a_1$, $a_2$. When plugged in the quantum mirror map for~$h_3$, we obtain the prediction of our exact quantization condition~(\ref{n3eqc}) for the spectrum of $\mH_3$. The calculation of the quantum mirror maps and the func\-tions~$\phi_{1,2}$ can be done in two ways: we can calculate them as a sum over gauge theory instantons, or we can calculate them by using the TBA equation. The instanton sum converges very slowly when $h_3$ is near the value
\begin{gather}\label{crit}
h_{3,{\rm c}}= 2 \Lambda^3 + 2\left( {h \over 3} \right)^{3/2}.
\end{gather}
The points $(h, h_3)$ satisfying this relation lie on the discriminant locus of the SW curve
\begin{gather*}
\Delta= \Lambda^{18} \big( 4 h^3 -27 \big(h_3+ 2 \Lambda^3\big)^2\big) \big( 4 h^3 -27 \big(h_3- 2 \Lambda^3\big)^2\big).
\end{gather*}
Since the instanton expansion in gauge theory is tailored for the region at infinity in the moduli space, it is only expected that it converges slowly near the discriminant locus. From the physical point view, at the critical value of $h_3$ given in (\ref{crit}), the regions where classical motion is allowed (the compact interval $[e_1^+, e_3^+]$ and the semi-infinite interval $(-\infty, e_2^+]$ depicted in Fig.~\ref{su3fig}) get together, and tunneling is no longer suppressed. This value marks then a transition between a~regime where resonances have small imaginary parts and behave almost as bound states, and a regime where resonances have real and imaginary parts of the same order. The TBA system can be solved by an iteration procedure which converges very well as long as $h_3<h_{3, {\rm c}}$, and leads to high numerical precision in this region.

In Tables \ref{tab:spectraN3-1} and \ref{tab:spectraN3-2} we compare the value of $h_3^{(n)}$ obtained with the numerical diagonalization, to the predictions of the quantization condition (\ref{n3eqc}) in two different situations. In evaluating the predictions, we use both the instanton sum and the TBA system when $h_3<h_{3, {\rm c}}$, while when $h_3>h_{3, {\rm c}}$ we use only the instanton sum. For the numerical diagonalization and the result obtained with the TBA method, we display the digits which are stable as we change the parameters that control the approximation. When using the instanton sum, we truncate the instanton sum in the quantum mirror map and in the NS free energy at the order indicated in the table, and we solve for the spectrum with this truncated sum. In the first example in Table~\ref{tab:spectraN3-1} we pick a value of $h= 5/2 (15/2)^{1/3}\approx 4.89{\dots}$ which is close to its value in the ground state of the~${\rm SU}(3)$ Toda lattice, and we focus on the ground state energy $h_3^{(0)}$. In Table~\ref{tab:spectraN3-2} we consider the value $h=10$ and we look at the levels $n=3$ and $n=6$. For the level with $n=6$, we evaluate the l.h.s.\ of the quantization condition~(\ref{n3eqc}) by using only the sum over gauge theory instantons. In all cases we have set $\hbar=\Lambda=1$.

\begin{table}[tb]\centering
 \begin{tabular}{ccc}\hline
	Instantons & $h_3^{(3)}$ & $h_3^{(6)}$ \\ \hline
	 $0$ & $10.031816233638+0.000665800161 \ri$ & $ \underline{24}.09065445717+\underline{8.5}522300579 \ri$ \\
	 $3$ & $\underline{9.9806278}40442+\underline{0.000764938}763\ri $ &$ \underline{24.1137974}5401+\underline{8.5338524}131 \ri$ \\
	 $5$ & $\underline{9.9806278228}05+\underline{0.000764938863} \ri $ & $\underline{24.1137974486}5+\underline{8.5338524066} \ri $ \\ \hline
	 TBA & $9.980627822889+0.000764938863 \ri $ &\\ \hline
	 numerical & $9.9806278228889+0.0007649388633 \ri $ & 24.1137974486+8.5338524066 \ri \\ \hline
 \end{tabular}
\caption{Resonant states of $\mH_3$ for $h=10$. We set $\hbar=\Lambda=1$.}\label{tab:spectraN3-2}
\end{table}

As we can see, the agreement is truly excellent, and leaves little room to doubt that~(\ref{n3eqc}) is indeed the correct quantization condition for this theory. We have of course performed many other tests,
and no discrepancy has been found between the numerical spectrum and the predictions of~(\ref{n3eqc}). We note that, for excited states, the determination of the
spectrum by using our quantization condition is numerically more efficient than Rayleigh--Ritz diagonalization.

\subsubsection[The case $N=4$]{The case $\boldsymbol{N=4}$}

We now look at the case $N=4$, where the potential is confining and it is given by
\begin{gather} \label{eqv4} V_4(x)=x^4 + x^2 h_2 - h_3 x .\end{gather}
The corresponding Hamiltonian is
\begin{gather*}
\mH_4 =\Lambda^4\big( \re^{\mm}+ \re^{-\mm} \big)+\mx^4 + h_2 \mx^2 -h_3 \mx .\end{gather*}
The vector \eqref{spec-vector} for $N=4$ is
\begin{gather*}
\bs{\gamma}= {1\over2} ( \bs{e}_1- \bs{e}_2+ \bs{e}_3- \bs{e}_4 ),
\end{gather*}
and its Weyl orbit contains six elements:
\begin{gather*}
\CW_3\cdot \bs{\gamma}= \bigg\{{1\over2} ( \bs{e}_1- \bs{e}_2+ \bs{e}_3- \bs{e}_4 ), {1\over2} ( \bs{e}_1+ \bs{e}_2- \bs{e}_3- \bs{e}_4), {1\over2} (\bs{e}_1- \bs{e}_2- \bs{e}_3+ \bs{e}_4), \\
\hphantom{\CW_3\cdot \bs{\gamma}= \bigg\{}{} {1\over2} ( -\bs{e}_1- \bs{e}_2+ \bs{e}_3+ \bs{e}_4),{1\over2}(-\bs{e}_1+ \bs{e}_2- \bs{e}_3+ \bs{e}_4),{1\over2}(-\bs{e}_1+ \bs{e}_2+ \bs{e}_3- \bs{e}_4)\bigg\}.
\end{gather*}
Therefore, the quantization condition \eqref{eqc-even} will contain six terms. If we define
\begin{gather*}
\phi_1={1\over \hbar}\left(-{\partial F_{\rm NS}\over \partial a_1}+2{\partial F_{\rm NS} \over \partial a_2} -{\partial F_{\rm NS} \over \partial a_3}\right) , \\
\phi_2= {1\over \hbar}\left(-2{\partial F_{\rm NS}\over \partial a_1}+{\partial F_{\rm NS} \over \partial a_2} \right) ,\qquad
\phi_3= {1\over \hbar}\left({\partial F_{\rm NS}\over \partial a_2}-2{\partial F_{\rm NS} \over \partial a_3} \right),
\end{gather*}
we can write it, explicitly, as
\begin{gather}
\big(\re^{\frac{2 \pi a_1}{\hbar }}-1\big) \big(\re^{\frac{2 \pi a_3}{\hbar }}-1\big) \re^{\frac{2 \pi a_2}{\hbar }-\ri (\phi_1-\phi_2-\phi_3)}
+\big(\re^{\frac{2 \pi a_1}{\hbar }}-1\big) \big(\re^{\frac{2 \pi a_3}{\hbar }}-1\big) \re^{\frac{2 \pi a_2}{\hbar }+\ri \phi_1}\nonumber\\
 \qquad{} +\re^{\ri (\phi_2+\phi_3)} \big(\re^{\frac{2 \pi (a_1+a_2)}{\hbar }}-1\big) \big(\re^{\frac{2 \pi (a_2+a_3)}{\hbar }}-1\big)+\re^{\ri \phi_2} \big(\re^{\frac{2 \pi a_2}{\hbar }}-1\big) \big(\re^{\frac{2 \pi (a_1+a_2+a_3)}{\hbar }}-1\big)\nonumber\\
\qquad{} +\re^{\ri \phi_3} \big(\re^{\frac{2 \pi a_2}{\hbar }}-1\big) \big(\re^{\frac{2 \pi (a_1+a_2+a_3)}{\hbar }}-1\big)+\big(\re^{\frac{2 \pi (a_1+a_2)}{\hbar }}-1\big) \big(\re^{\frac{2 \pi (a_2+a_3)}{\hbar }}-1\big)=0. \label{n4ecq}
 \end{gather}
On the other hand, the quantization conditions \eqref{toda-qc} for the ${\rm SU}(4)$ Toda lattice are given by
\begin{gather}\label{toda4-qc}\re^{\ri \phi_i}=-1, \qquad i=1,2,3 .\end{gather}
It is also useful to introduce explicit quantum numbers as in \eqref{toda-qc}, and to write \eqref{toda4-qc} as
\begin{gather}\label{toda4-qc2} \phi_1= 2 \pi \left(\ell_2+{1\over 2}\right) , \qquad -\phi_2= 2 \pi \left(\ell_1+{1\over 2}\right), \qquad -\phi_3=2 \pi \left(\ell_3+{1\over 2}\right) .\end{gather}
These are the same quantum numbers appearing in (\ref{4qn}). It is easy to verify that when \eqref{toda4-qc} holds, then \eqref{n4ecq} holds as well.

A particularly interesting case occurs when $a_1=a_3$, hence $h_3=0$. In this case, the Hamiltonian $\mH_4$ is of the form (\ref{H40}), and the eigenstates have a definite parity. The quantization condition \eqref{n4ecq} can be simplified to the form
\begin{gather}\label{symqc}
\big(\re^{\ri \phi_2/2}+\re^{-\ri \phi_2/2} \big)^2= 4 \CV^2 \sin^2 \left( {\phi_1 - \phi_2 \over 2} \right),
\end{gather}
where
\begin{gather*}
\CV= \re^{- \pi a_2/\hbar} \frac{1-\re^{-2 \pi a_1/\hbar }}{1-\re^{-2 \pi (a_1+a_2)/\hbar }}.
\end{gather*}
We note that the r.h.s.\ of (\ref{symqc}) is purely non-perturbative if $a_i>0$. As in the conventional double-well potential, we expect that extracting the square root of~(\ref{symqc}) will introduce explicitly a sign keeping track of the parity of the states. Therefore, we write
\begin{gather}\label{symqc2}
\re^{\ri \phi_2/2}+\re^{-\ri {\phi}_2/2}= 2 \epsilon \CV \sin \left( {\phi_1 - \phi_2 \over 2} \right), \qquad \epsilon=\pm 1.
\end{gather}
We indeed find that the energy levels $E_{2k}$, $E_{2k+1}$ correspond to $\epsilon=\pm (-1)^k$, respectively. Moreover, the r.h.s.\ of (\ref{symqc2}) vanishes at the Toda lattice points \eqref{toda4-qc2}. This indicates that two successive eigenstates of opposite parity are degenerate, in agreement with the results presented at the end of Section~\ref{sec:sp}.

\begin{figure}[t!]\centering
\includegraphics[scale=0.6]{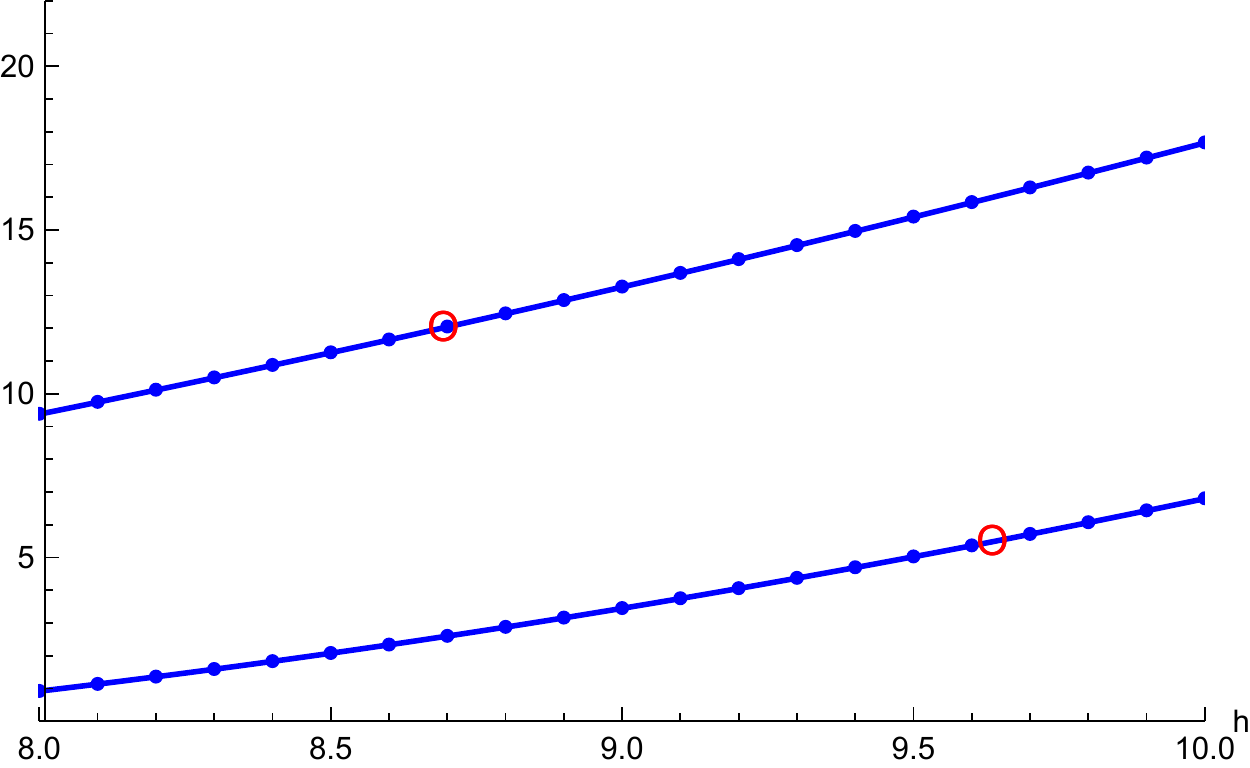}
\caption{The function $h_4^{(n)}(h)$ as given in equation \eqref{hh4} for $n=1$ (lower line) and $n=2$ (upper line). The dots are obtained from the quantization condition \eqref{n4ecq}, while the continuous lines by using numerical diagonalization. We set $\Lambda=\hbar=1$. The red circles indicate two Toda lattice points.}\label{E23}
\end{figure}

We now give numerical evidence for the exact quantization condition \eqref{n4ecq}. We proceed as in the case $N=3$, namely, we fix the coefficients $h_2$ and $h_3$ in \eqref{eqv4} to given values, and we calculate the eigenvalues of $\mH_4$ by numerical diagonalization. These numerical results are then compared to the spectrum obtained from \eqref{n4ecq}, which is obtained by solving \eqref{n4ecq} together with the condition that~$h_2$,~$h_3$, as a function of the $a_i$'s, are fixed:
\begin{gather*}
h_2 = h_2(a_1,a_2,a_3, \hbar), \qquad h_3=h_3(a_1,a_2,a_3, \hbar).
\end{gather*}
In this way we produce a discrete set of values for the $a_i$'s
\begin{gather*} \big\{ a_1^{(n)},a_2^{(n)},a_3^{(n)}\big\}_{n\geq 0},\end{gather*}
which we plug into the mirror map for $h_4$ to obtain
\begin{gather} \label{h4sqc} h_4^{ (n)}= h_4\big(a_1^{(n)},a_2^{(n)},a_3^{(n)}, \hbar\big). \end{gather}
We can then check that the values obtained from \eqref{h4sqc} match the values obtained by numerical diagonalization of $\mH_4$. An example is given in Table \ref{tab:spectraN4-1}. In a similar way we also tested~\eqref{symqc2}. An example is given in Table~\ref{tab:spectraN4-2}. When $h_3=0$ the quantization condition~\eqref{n4ecq} leads to a function
\begin{gather}\label{hh4} h_4^{(n)}(h), \qquad n=0,1,2,\dots, \end{gather}
where we recall that $h=-h_2$. This function is shown in Fig.~\ref{E23} for $n=1$ and $n=2$. We emphasize that our quantization condition leads to a discrete family of codimension one submanifolds in moduli space, as it is clear from Fig.~\ref{E23}. This is in contrast to the solutions to the Baxter equation, which lead to a discrete set of points in moduli space.

As a final remark we note that, similarly to what we discussed for the $N=3$ potential, also when $N=4$ there are values of the energy which are deep inside moduli space, far from the semiclassical region at infinity, and for which the instanton sum converges very slowly.

\begin{table}[t]\centering
 \begin{tabular}{cc}\hline
	Instantons & $h^{(0)}_4$ \\ \hline
	 $1$ & $\underline{4.760}407026825$\\
	 $2$ & $ \underline{4.76029}1739919$ \\
	 $3$ & $ \underline{4.7602900}29259$ \\
	 $4$ & $\underline{4.76029001214}6$ \\ \hline
	 TBA & \qquad $ 4.7602900121484067$ \\ \hline
	 numerical & \qquad $ 4.7602900121484067$ \\ \hline
 \end{tabular}
\caption{Ground state energy $E_0=-h^{(0)}_4$ of $\mH_4$ for $h_2=-6$ and $h_3=1$. We set $\hbar=\Lambda=1$.}\label{tab:spectraN4-1}
\end{table}

\begin{table}[t!]\centering
 \begin{tabular}{ccc}\hline
	Instantons & $h^{(0)}_4$ & $h^{(1)}_4$ \\ \hline
	 $1$ &$\underline{17.678}46191066723095$ & $ \underline{17.67}800109302647040$\\
	 $2$ &$ \underline{17.6784423}6054395243$ &$\underline{17.6779815}4085570493$ \\
	 $3$ &$\underline{17.6784423707}3924490$ &$\underline{17.6779815510}4316065$\\
	 $4$ &$\underline{17.678442370757}58180$ &$ \underline{17.6779815510614}9680$ \\\hline
	 TBA & $17.67844237075748709 $ & $17.67798155106140212 $ \\ \hline
	 numerical& $17.67844237075748709 $ & $17.67798155106140212$ \\ \hline
 \end{tabular}
\caption{The first two energy levels $E_{n}=-h^{(n)}_4$ of $\mH_4$ for $h_2=-10$ and $h_3=0$, as obtained from \eqref{symqc2} with $\epsilon=1$ and $\epsilon=-1$, respectively. We set $\hbar=\Lambda=1$.}\label{tab:spectraN4-2}
\end{table}

 \section{Derivation from the TS/ST correspondence}\label{derivation}

\subsection{The TS/ST correspondence}
We now summarize the main ingredients of the TS/ST correspondence in the higher genus case. Our presentation is very synthetic and it mainly serves the goal of setting our notation. We refer the reader to \cite{cgm,cgum,mmrev} for more details and background.

Toric CY threefolds are specified by a matrix of charges $Q_i^\alpha$, $i=0, \dots, k+2$, $\alpha=1, \dots, k$, satisfying the condition,
\begin{gather*}
\sum_{i=0}^{k+2} Q_i^\alpha=0, \qquad \alpha = 1, \ldots, k . 
\end{gather*}
The Newton polygon $\CN$ associated to these charges is given by a set of points $\bs{\nu}^i=(\nu_1^i, \nu_2^i)$, $i=0,1, \dots, k+2$, satisfying the constraints
\begin{gather*}
\sum_{i=0}^{k+2} Q_i^\alpha \bs{\nu}^i=0, \qquad \alpha=1, \dots, k.
\end{gather*}
The mirror curve to the toric CY is described by the curve,
\begin{gather*}
W(\re^x, \re^p)=0,
\end{gather*}
where
\begin{gather}\label{wmc}
W(\re^x, \re^p)= \sum_{i=0}^{k+2} a_i \re^{\nu_1^i x + \nu_2^i p}.
\end{gather}
The coefficients $a_i$, $i=0, \dots, k+2$, are not independent, and we should use instead the Batyrev coordinates
\begin{gather*}
z_\alpha= \prod_{i =0}^{k+2} a_i^{Q_i^\alpha}, \qquad \alpha=1, \dots, k.
\end{gather*}
The curve (\ref{wmc}) can be quantized by promoting $x$, $p$ to Heisenberg operators $\mx$, $\mm$ satisfying the canonical commutation relation
\begin{gather*}
[\mx, \mm]=\ri \hbar
\end{gather*}
and using Weyl's ordering prescription. We want to solve the eigenvalue problem
\begin{gather}\label{qmc-spec}
W\big(\re^\mx, \re^\mm\big) |\psi \rangle=0.
\end{gather}
The quantization condition determining the spectrum is a single constraint among the values of the Batyrev coordinates, as emphasized in~\cite{cgm}. According to the conjecture put forward in~\cite{cgm,ghm}, this constraint can be obtained by using information from topological string theory on the CY threefold. The information we need is the following. First, we need the Gopakumar--Vafa free energy~\cite{gv}
\begin{gather*}
F_{\rm GV} (\bs{t}, g_s )=\sum_{g\ge 0} \sum_{\bs{d}} \sum_{w=1}^\infty {1\over w} n_g^{ \bs{d} } \left(2 \sin { w g_s \over 2} \right)^{2g-2} \re^{-w \bs{d} \cdot \bs{t}}.
\end{gather*}
In this equation, $\bs{ t}=(t_1, \dots, t_k)$ is the vector of complexified K\"ahler parameters, $g_s$ is the string coupling constant, and $n_g^{\bf d}$ are the Gopakumar--Vafa invariants. The topological string free energy has also a perturbative piece, which we write as
\begin{gather}\label{pertf}
F_{\rm p}(\bs{t}, g_s)= \frac{1}{6 g_s^2}\sum_{i,j,k=1}^{k} a_{ijk} t_i t_jt_k + \sum_{i=1}^{k} \left(b_i + \frac{4\pi^2}{g_s^2} b_i^{\rm NS}\right)t_i.
\end{gather}
Here, $a_{ijk}$, $b_i$, $b_i^{\rm NS}$ are coefficients which can be explicitly computed for any toric CY model (see~\cite{hkp} for examples). In addition, we need the NS free energy of the toric CY, $F^{\rm BPS}_{\rm NS}(\bs{t},\hbar)$, which can be expressed in terms of the refined BPS invariants $N^{\bs{d}}_{j_L, j_R} $ of~\cite{ikv} as
\begin{gather*}
F^{\rm BPS}_{\rm NS}(\bs{t}, \hbar) = \sum_{j_L, j_R} \sum_{w, \bs{d} }N^{\bs{d}}_{j_L, j_R} \frac{\sin\frac{\hbar w}{2}(2j_L+1)\sin\frac{\hbar w}{2}(2j_R+1)}{2 w^2 \sin^3\frac{\hbar w}{2}} \re^{-w \bs{ d}\cdot\bs{ t}} .
\end{gather*}
Another ingredient we need is the $\bs{B}$ field first introduced in \cite{hmmo}. This is a $k$-component vector which satisfies the following requirement: for all ${\bf d}$, $j_L$ and $j_R$ such that $N^{{\bf d}}_{j_L, j_R} $ is non-vanishing, we must have
\begin{gather*}
(-1)^{2j_L + 2 j_R+1}= (-1)^{{\bs {B}} \cdot {\bf d}}.
\end{gather*}
Finally, we need an additional fact about local CY manifolds. In these manifolds, the moduli are of two types: we have $g_\Sigma$ moduli
\begin{gather*}
\kappa_i=\re^{\mu_i}, \qquad i=1, \dots, g_\Sigma,
\end{gather*}
and in addition we have ``mass parameters'' $\xi_j$, $j=1, \dots, r_\Sigma$, where $r_\Sigma=k-g_\Sigma$. The Batyrev coordinates can be written as
\begin{gather*}
-\log z_i= \sum_{j=1}^{g_\Sigma}C_{ij} \mu_j + \sum_{j=1}^{r_\Sigma} \alpha_{ij}\log {\xi_j}, \qquad i=1, \dots, k.
\end{gather*}
We now define the so-called modified grand potential of the CY $X$ as
\begin{gather*}
\mJ_X(\bs{\mu},\bs{\xi},\hbar) = F_{\rm p}\left( \frac{2\pi}{\hbar}\bs{t}(\hbar), \frac{4\pi^2}{\hbar} \right)+ \sum_{i=1}^{k}\frac{t_i(\hbar)}{2\pi}\frac{\partial}{\partial t_i} F^{\rm BPS}_{\rm NS}(\bs{t}(\hbar),\hbar) + \frac{\hbar^2}{2\pi}\frac{\partial}{\partial \hbar}\left( \frac{F^{\rm BPS}_{\rm NS}(\bs{t}(\hbar),\hbar)}{\hbar} \right)\nonumber\\
\hphantom{\mJ_X(\bs{\mu},\bs{\xi},\hbar) =}{} + F_{\rm GV}\left( \frac{2\pi}{\hbar}\bs{t}(\hbar) + \pi \ri \bs{B}, \frac{4\pi^2}{\hbar} \right).
\end{gather*}
In this equation, $\bs{t}(\hbar)$ is the quantum mirror map \cite{acdkv}, which is a function of $\hbar$ and the Batyrev coordinates (therefore of the moduli and the mass parameters). We recall that, in the large radius region of moduli space,
\begin{gather*}
t_i (\hbar)= -\log z_i + \CO( z_k).
\end{gather*}

We are now ready to define the quantum theta function of the model, as
\begin{gather*}
\Theta_X({\boldsymbol \kappa}; \hbar)= \sum_{ {\bf m} \in \IZ^{g_\Sigma}} \exp\big( \mathsf{J}_{X}(\boldsymbol{\mu}+2 \pi \ri {\bf m}, \boldsymbol{\xi}, \hbar) -\mathsf{J}_{X}(\boldsymbol{\mu}, \boldsymbol{\xi}, \hbar)\big).
\end{gather*}
It was conjectured in \cite{cgm,ghm} that the vanishing of this quantum theta function gives the quantization condition for the spectral problem~(\ref{qmc-spec}). This conjecture has been verified in many examples. In particular, in the higher genus case, it has been checked in \cite{bgt2, cgm,cgum,swh}. As we will now show, our quantization conditions (\ref{eqc-odd}), (\ref{eqc-even}) are limiting cases of this conjecture for a~particular family of toric CY manifolds.

 \subsection{The 4d limit}

 The geometries we will consider are the toric CY known in the literature as resolved $Y^{N,0}$ singularities (see, e.g.,~\cite{bt,hm} for a useful summary of their properties). These geometries engineer 4d SW theory with gauge group ${\rm SU}(N)$ in an appropriate limit~\cite{kkv}. Their mirror curve is precisely the SW curve of 5d, ${\rm SU}(N)$ supersymmetric Yang--Mills theory (\ref{rt-sc}), which we will write as
\begin{gather}\label{mcurve}
W\big(\re^x, \re^p\big)=a_1 \re^p + a_2 \re^{-p}+ \sum_{i=0}^N \kappa_i \re^{(N/2-i) x}=0.
\end{gather}
The coefficients appearing here are related to the ones in (\ref{rt-sc}) and (\ref{rt-sc-N}) by
\begin{gather*}
a_1=a_2 = \left( R \Lambda \right)^N, \qquad \kappa_i=(-1)^i H_i, \qquad i=0, 1, \dots, N,
\end{gather*}
and we recall that $H_0=H_N=1$. This CY has $N-1$ moduli, which can be taken to be the coefficients $\kappa_i$, $i=1, \dots, N-1$, and one mass parameter, which is $\Lambda R$. The Batyrev coordinates are
\begin{gather*}
z_i= {\kappa_{i-1} \kappa_{i+1} \over \kappa^2_i}={H_{i-1} H_{i+1} \over H_i^2}, \qquad i=1, \dots, N-1, \nonumber\\
 z_N ={a_1 a_2 \over \kappa_0 \kappa_N}= (-1)^N (R \Lambda)^{2N}.
\end{gather*}
The model has $N$ K\"ahler moduli $t_i$, $i=1, \dots, N$. The first $N-1$ correspond to true moduli, and their quantum mirror map is given by
\begin{gather*}
t_i (\hbar)=\sum_{j=1}^{N-1} C_{ij} \mu_j +\CO\left(\re^{\mu_k} \right), \qquad i=1, \dots, N-1,
\end{gather*}
where $C_{ij}$ is the Cartan matrix of ${\rm SU}(N)$, while
\begin{gather}\label{tnzn}
t_N = -\log z_N
\end{gather}
is a parameter. We note that the relation (\ref{tnzn}) has no corrections. As in the previous section, we will collect the ``true'' K\"ahler moduli $t_1, \dots, t_{N-1}$ in a vector of $\Lambda_{\rm w} \otimes \IC$,
\begin{gather*}
\bs{t}=\sum_{j=1}^{N-1} t_j \bs{\lambda}_j.
\end{gather*}
This should not be confused with the generic vector $\bs{t}$ appearing the formulae of the previous subsection. The modified grand potential of the geometry can be computed by using standard tools in topological string theory. The perturbative free energy (\ref{pertf}) is given by
\begin{gather}\label{f0p}
F_{\rm p} (\bs{t}, t_N, g_s)= {1\over 6g_s^2} \sum_{\balpha \in \Delta_+} (\bs{t}\cdot \balpha)^3 + {t_N \over 2 N g_s^2 } \sum_{\balpha \in \Delta_+} (\bs{t}\cdot \balpha)^2 +
{1\over 6}\left( 1-{4 \pi^2 \over g_s^2} \right) \bs{t} \cdot \bs{\rho}.
\end{gather}
The GV and the NS free energies are of the form
\begin{gather}
F_{\rm GV}( \bs{t}, t_N, g_s) = \CF_{\rm GV}(\bs{t}, g_s)+\CO\big(\re^{-t_N} \big), \nonumber\\
 F^{\rm BPS}_{\rm NS} ( \bs{t}, t_N, \hbar) = \CF_{\rm NS} (\bs{t}, \hbar) + F_{\rm NS}^{\rm 5d}(\bs{Q}, \hbar),\label{bps-piece}
\end{gather}
where
\begin{gather}
\CF_{\rm NS}(\bs{t}, \hbar) =-\sum_{\balpha \in \Delta_+} \sum_{w\ge 1} {1\over w^2} \cot\left( {\hbar w \over 2}\right) \re^{-w \balpha\cdot \bs{t}}, \nonumber\\
 \CF^{\rm GV} (\bs{t},g_s) =-2 \sum_{\balpha \in \Delta_+}\sum_{v \ge 1} {1\over v} {1\over 4 \sin^2\left( {g_s v \over 2} \right)} \re^{- v \balpha\cdot \bs{t}},\label{FF}
\end{gather}
do not depend on $t_N$. In the second line of (\ref{bps-piece}), $F_{\rm NS}^{\rm 5d}(\bs{Q}, \hbar)$ is the 5d NS free energy defined in~(\ref{5dns}), after the identification
\begin{gather*}
\bs{t}=\sum_{I=1}^N \alpha_I \bs{e}_I.
\end{gather*}
The $\bs{B}$-field vanishes when $N$ is even, and it is given by
\begin{gather*}
\bs{B}=(0, \dots, 0,1)
\end{gather*}
when $N$ is odd, but will not be relevant for the 4d limit we are considering. Finally, we note that in the calculation of the quantum theta function, the shift
\begin{gather*}
\mu_j \rightarrow \mu_j+ 2 \pi \ri m_j, \qquad j=1, \dots, N-1,
\end{gather*}
where the $m_j$ are integers, corresponds to the following shift in the K\"ahler moduli,
\begin{gather*}
t_j \rightarrow t_j+ 2 \pi \ri \sum_k C_{jk} m_k.
\end{gather*}
In terms of vectors in $\Lambda_{\rm w} \otimes \IC$, this can be written as
\begin{gather}\label{tm-shift}
\bs{t} \rightarrow \bs{t}+2 \pi \ri \bs{m}, \qquad \bs{m}= \sum_{j=1}^{N-1} m_j \balpha_j \in \Lambda_{\rm r},
\end{gather}
where $\Lambda_{\rm r}$ is the root lattice. However, as illustrated in \cite{cgm,ghm}, one should re-express the moduli appearing in the curve in terms of the natural spectral quantities. For example, in curves of genus one, the modulus $\kappa$ is related to the energy by $\kappa=-E$. In our case, the natural moduli are the Hamiltonians $H_i$, which for $i$ odd differ from the $\kappa_i$ in a sign. Therefore, we should redefine
\begin{gather}\label{mu-shift}
\mu_i\rightarrow \mu_i + \pi \ri, \qquad \text{$i$ odd}, \qquad 1\le i \le N-1.
\end{gather}
In addition, when $N$ is odd, $\kappa_N=-1$, and this leads to an extra shift of $-\pi \ri$ in $t_{N-1}$. It easy to see that these additional shifts can be incorporated in the quantum theta function if we further shift $\bs{t}$ by
\begin{gather*}
\bs{t} \rightarrow \bs{t}+ 2 \pi \ri \bs{\gamma},
\end{gather*}
where $\bs{\gamma} \in \Lambda_w$ has been defined in (\ref{spec-vector}). To see this, we note that, when $N$ is even,
\begin{gather*}
\bs{\gamma}= {1\over 2} ( \bs{\alpha}_1+ \bs{\alpha}_3+\cdots+ \bs{\alpha}_{N-1} ),
\end{gather*}
and leads to a shift by $\pi \ri $ in $\mu_i$ when $i$ is odd. This is precisely what is needed in~(\ref{mu-shift}). When~$N$ is odd, we can write
\begin{gather*}
\bs{\gamma}= {1\over 2} ( \bs{\alpha}_1+ \bs{\alpha}_3+\cdots+ \bs{\alpha}_{N-2}- \bs{\lambda}_{N-1} ).
\end{gather*}
The simple roots lead to the shift (\ref{mu-shift}), while the last term in the r.h.s.\ leads to an additional shift of $-\pi \ri $ in $t_{N-1}$. As a consequence, the vector $\bs{m}$ in~(\ref{tm-shift}) has to be replaced by
\begin{gather*}
\bs{n} = \bs{m} + \bs{\gamma}.
\end{gather*}

The spectral problem corresponding to the curve (\ref{mcurve}) can be studied in detail on its own (the case $N=3$ was analyzed in~\cite{cgum} while $N>3$ was studied in~\cite{bgt2}). However, we are here interested in the spectral curve~(\ref{curve}), which is obtained in the 4d limit~(\ref{R4d}). In particular, the Planck constant associated to the quantization of~(\ref{mcurve}), which we will denote by $\hbar_{\rm 5d}$ from now on, has to scale as
\begin{gather}\label{h5d}
\hbar_{\rm 5d}= R \hbar,
\end{gather}
where $\hbar$ will be identified with the Planck constant appearing in (\ref{4dpc}). In addition, we have the scaling
\begin{gather*}
t_i= R a_i, \qquad i=1, \dots, N-1.
\end{gather*}
This relates the moduli of the CY to the moduli of the SW curve. The exponentiated K\"ahler modulus
\begin{gather}\label{tnsup}
\re^{-t_N}= (-1)^N \left( R \Lambda \right)^{2N},
\end{gather}
as well as the ``dual" modulus
\begin{gather*}
\re^{-2 \pi t_N/\hbar_{\rm 5d}}
\end{gather*}
vanish when $R\rightarrow 0$. We note that this limit is in some sense opposite to the one considered in~\cite{bgt,bgt2}, where $\hbar_{\rm 5d} \to \infty$ and only the ``dual'' moduli survive.

It turns out that, in the 4d limit (\ref{R4d}), the quantum theta function collapses to a {\it finite sum}. To understand why this is the case, we note that the quantum theta function is an infinite sum of terms. Each term is associated to an element in the root lattice $\bs{m} \in \Lambda_r$ and contains a factor of the form
\begin{gather}\label{sup-factor}
\exp\left( -{2 \pi N \over \hbar R} \log\left({1\over R \Lambda}\right) \big( \bs{n}^2-\bs{\gamma}^2 \big) \right),
\end{gather}
where $\bs{n}= \bs{m}+ \bs{\gamma}$. This factor comes from the second term in~(\ref{f0p}), after using the identity
\begin{gather*}
\sum_{\balpha \in \Delta_+} (\bs{v}\cdot \balpha)(\balpha\cdot \bs{w})= N \bs{v}\cdot \bs{w}.
\end{gather*}
It turns out that $ (\bs{m}+ \bs{\gamma} )^2- \bs{\gamma}^2 \ge 0$ for any $\bs{m} \in \Lambda_r$.\footnote{Although we do not have a rigorous proof of this property, we have checked it extensively. In general, given $\bs{w}\in \Lambda_w$, it is not true that $\left(\bs{m}+ \bs{w}\right)^2- \bs{w}^2 \ge 0$ for all $\bs{m} \in \Lambda_r$. This seems to be valid only for weights which belong to the Weyl orbit of a fundamental weight. This is the case for $\bs{\gamma}$, which is in the Weyl orbit of $\bs{\lambda}_{(N+1)/2}$ when $N$ is odd, and in the Weyl orbit of $\bs{\lambda}_{N/2}$ when $N$ is even.} Therefore,~(\ref{sup-factor}) leads to an exponential suppression unless
\begin{gather*}
\bs{n}^2= \bs{\gamma}^2.
\end{gather*}
Due to the Weyl invariance of the Cartan--Killing form, this equality holds when $\bs{n}$ is an element in the Weyl orbit of~$\bs{\gamma}$. Therefore, terms in the quantum theta function for which $\bs{n} \notin \CW_N \cdot \bs{\gamma}$ simply vanish in the 4d limit.

The basic ingredient in the quantum theta function is the modified grand potential $\mJ_X$. In the GV contribution to~$\mJ_X$, terms proportional to $\re^{-t_N}$ vanish in the 4d limit, as a consequence of~(\ref{tnsup}), and only $\CF_{\rm GV}$ in (\ref{FF}) survives. This function should be combined with the function~$\CF_{\rm NS}$ appearing in the same equation~(\ref{FF}). To evaluate their contribution, we consider the function
\begin{gather*}
f(Q, \hbar)=\sum_{w \ge 1} \left\{ {1\over w} \cot\left( {\hbar w \over 2} \right) Q^w+ {1\over w} \cot\left( {\hbar_D w \over 2} \right) Q_D^w\right\},
\end{gather*}
where
\begin{gather*}
\hbar_D= {4 \pi^2 \over \hbar}.
\end{gather*}
This function was introduced in \cite{hm} in order to understand the 4d limit of the quantization conditions for the relativistic Toda lattice. It has the following integral representation,
\begin{gather*}
f(Q, \hbar)= \frac{2}{\hbar} {\rm Li}_2(Q)+\frac{2}{\pi} {\rm Re} \left[ \int_0^{\infty \re^{\pm \ri 0}}\frac{\hbar Q(\cosh \hbar x-Q)}{(1-Q \re^{\hbar x})(1-Q \re^{-\hbar x})}\log\big(1-\re^{-2\pi x}\big) \rd x \right] .
\end{gather*}
In the 4d limit (\ref{R4d}) we set,
\begin{gather*}
Q =\re^{-R a},
\end{gather*}
and one has \cite{hm}
\begin{gather}\label{f4d}
f(Q, \hbar_{\rm 5d})= {\pi^2 \over 3 \hbar R} + {2\over \hbar} a \log ( R \Lambda ) + {2\over \hbar} \gamma_{\rm 4d}(a, \hbar) +\CO(R),
\end{gather}
where $\hbar_{\rm 5d}$ is given in (\ref{h5d}) and $\gamma_{\rm 4d} (a, \hbar)$ was defined in (\ref{gam4d}).

Let us now calculate in detail the 4d limit of $\mJ_X(\bs{\mu}+ 2 \pi \ri \bs{n}, z_N, \hbar)-\mJ_X(\bs{\mu}, z_N, \hbar)$ when $\bs{n}$ belongs to the Weyl orbit~(\ref{orbit}). The contribution of the functions in~(\ref{FF}) is
\begin{gather}\label{fnp}
\ri \sum_{\balpha \in \Delta_+} (\balpha\cdot \bn) \big( f(Q_{\balpha}, \hbar_{\rm 5d}) + \ri (\balpha\cdot \bn) \log (1- Q_{\balpha,D} )\big),
\end{gather}
where
\begin{gather*}
Q_{\balpha}=\re^{- \balpha\cdot \bs{t}}, \qquad Q_{\balpha,D}= \re^{-2 \pi \balpha\cdot \bs{t}/ \hbar_{\rm 5d} }.
\end{gather*}
In deriving this result, we have taken into account that, when $\bs{n}$ belongs to the Weyl orbit~(\ref{orbit}) and $\balpha \in \Delta_+$, the possible values of $\bs{n}\cdot \balpha$ are~$0$,~$\pm1$, so that
\begin{gather*}\begin{split}&
 \sin \left( {2 \pi^2 v \over \hbar_{\rm 5d} } (\bs{n}\cdot \balpha) \right)\re^{-2 \pi^2 \ri v (\bs{n}\cdot \balpha)/ \hbar_{\rm 5d}}\\
& \qquad {} = (\bs{n}\cdot \balpha) \sin \left( {2 \pi^2 v \over \hbar_{\rm 5d}} \right)\left\{ \cos \left( {2 \pi^2 v \over \hbar_{\rm 5d}} \right)
 -\ri (\bs{n}\cdot \balpha) \sin \left( {2 \pi^2 v \over \hbar_{\rm 5d}} \right) \right\}.\end{split}
\end{gather*}
The term involving $F_{\rm NS}^{\rm 5d}$ combines with the terms involving $\gamma_{\rm 4d}$ in (\ref{fnp}) to give a contribution
\begin{gather*}
{\ri \over \hbar} \bs{n}\cdot {\partial F_{\rm NS} \over \partial \bs{a}}.
\end{gather*}
 From the first term in the r.h.s.\ of~(\ref{f0p}) we obtain
\begin{gather*}
-{\pi \over \hbar} \sum_{\balpha \in \Delta_+} \left(\bs{n}\cdot \balpha\right)^2 (\bs{a}\cdot \balpha).
\end{gather*}
This combines with the limit of the second term in~(\ref{fnp})
\begin{gather*}
-\sum_{\balpha \in \Delta_+} (\bs{n}\cdot \balpha )^2 \log\big(1- \re^{-2 \pi \bs{a}\cdot \balpha/\hbar} \big).
\end{gather*}
into
\begin{gather*}
-\sum_{\balpha \in \Delta_+} (\bs{n}\cdot \balpha )^2 \log\left[ 2 \sinh\left( {\pi \bs{a}\cdot \balpha\over \hbar}\right)\right].
\end{gather*}
When $N$ is odd, there is also a contribution coming from $t_{N}= \pi \ri+\cdots$ in the second term of~(\ref{f0p}), which gives
\begin{gather*}
-{\pi \over N \hbar} \sum_{\balpha \in \Delta_+} (\bs{n}\cdot \balpha ) (\bs{a}\cdot \balpha)= -{\pi \over \hbar} \bs{n}\cdot \bs{a}.
\end{gather*}

We can also check that terms which diverge as $R \rightarrow 0$ cancel (similar cancellations occur in~\cite{hm}). Let us first consider terms proportional to~$1/R$. The first term in (\ref{f0p}) gives
\begin{gather*}
-{2 \pi^2 \ri \over 3 \hbar R} \sum_{\alpha \in \Delta_+} (\bn \cdot \balpha)^3= -{4 \pi^2 \ri \over 3 \hbar R} \bn \cdot \bs{\rho}.
\end{gather*}
The last term in (\ref{f0p}) leads to a divergent shift
\begin{gather*}
 {2 \pi^2 \ri \over 3 \hbar R} \bn \cdot \bs{\rho}.
\end{gather*}
Finally, the first term in (\ref{fnp}) gives, after using~(\ref{f4d}),
\begin{gather*}
{ \pi^2 \ri \over 3 \hbar R} \sum_{\alpha \in \Delta_+} \bn \cdot \balpha={2 \pi^2 \ri \over 3 \hbar R} \bn \cdot \bs{\rho},
\end{gather*}
so that everything cancels. Since
\begin{gather*}
t_{N}= -2 N \log\left(R \Lambda \right)+\cdots
\end{gather*}
the second term in (\ref{f0p}) also gives a logarithmically divergent term. This cancels against the logarithmic divergences appearing in the 4d limit of the first term in~(\ref{fnp}).

We conclude that, in the 4d limit, $\mJ_X(\bs{\mu}+ 2 \pi \ri \bs{n}, z_N, \hbar)-\mJ_X(\bs{\mu}+ 2 \pi \ri \bs{\gamma}, z_N, \hbar)$ becomes
\begin{gather*}
 {\ri \over \hbar} {\partial F_{\rm NS} \over \partial \bs{a}} \cdot (\bs{n}-\bs{\gamma})
 +\sum_{\balpha \in \Delta_+}\big( (\bs{\gamma}\cdot \balpha )^2- (\bs{n}\cdot \balpha )^2\big) \log\left[ 2 \sinh\left( {\pi \bs{a}\cdot \balpha\over \hbar}\right)\right]
 \end{gather*}
 when $N$ is even, and
\begin{gather*}
 {\ri \over \hbar} \left( {\partial F_{\rm NS} \over \partial \bs{a}}+ \pi \ri \bs{a} \right) \cdot ( \bs{n}- \bs{\gamma} )
 +\sum_{\balpha \in \Delta_+}\big( (\bs{\gamma}\cdot \balpha )^2- (\bs{n}\cdot \balpha )^2 \big)\log\left[ 2 \sinh\left( {\pi \bs{a}\cdot \balpha\over \hbar}\right)\right]
\end{gather*}
when $N$ is odd. Since the quantization condition is given by the vanishing of the quantum theta function, we can factor out the terms involving~$\bs{\gamma}$. We recover in this way our exact quantization conditions~(\ref{eqc-even}), (\ref{eqc-odd}) from the TS/ST correspondence.

It is interesting to note that the non-perturbative corrections in (\ref{eqc-even}), (\ref{eqc-odd}) involving the periods $\pi a_i$ come from the last term in (\ref{fnp}). This term has its origin in the contribution of the {\it standard} topological string free energy to $\mJ_X$. In \cite{cgm,ghm, km} it was shown that the contribution of the standard topological string is essential for the exact solution of quantum mirror curves. Our derivation shows that this contribution is also crucial for the quantization of the SW curve. Interestingly, such contributions do not play a r\^ole in the quantization conditions for the Toda lattice (\ref{toda-qc}), which only involve the all-orders WKB/EBK periods, i.e., the quantum $B$-periods, and can be regarded as purely perturbative.

\section{The limit of standard quantum mechanics}\label{sec-qmrel}

The Hamiltonian (\ref{dqmh}) is clearly a deformation of the standard Hamiltonian (\ref{sqmh}). We would like to make this statement more precise, since we would eventually like to use our exactly solvable deformation to shed light on conventional quantum mechanics. It turns out that the conventional Hamiltonian~(\ref{sqmh}) emerges in a very non-trivial regime of the underlying supersymmetric gauge theory: we have to approach the AD superconformal point~\cite{ad} inside the moduli space of the~${\rm SU}(N)$ theory~\cite{ehiy}.

Let us see this in detail. It is clear that, in order to recover (\ref{sqmh}), we should consider small values of the momentum. Let us introduce a scaling parameter $\alpha$, which we will eventually take to zero, and let us consider the following scaling
\begin{gather*}
\mm = \alpha^{\lambda_1} \widetilde \mm, \qquad \mx = \alpha^{\lambda_2} \widetilde \mx,
\end{gather*}
so that
\begin{gather}\label{ll12}
\lambda_1>0, \qquad \lambda_2>0, \qquad \lambda_1+\lambda_2=1,
\end{gather}
The values of the exponents $\lambda_{1,2}$ will be fixed in a moment. We will also scale the Planck constant as
\begin{gather*}
\hbar= \alpha \hbar_{\rm QM},
\end{gather*}
so that
\begin{gather*}
[\widetilde \mm, \widetilde \mx]= \hbar_{\rm QM}.
\end{gather*}
We now scale the coefficients in the potential
\begin{gather*}
W_N(\mx)= \sum_{k =0}^{N-1} (-1)^k \mx^{N-k} h_k + (-1)^N h_N
\end{gather*}
as
\begin{gather*}
h_k = \alpha^{\Delta_k} \widetilde h_k, \qquad k=2, \dots, N-1,
\end{gather*}
and
\begin{gather*}
h_N= 2(-1)^{N-1}\Lambda^N+ \alpha^{\Delta_N} \widetilde h_N.
\end{gather*}
We then find
\begin{gather*}
\mH_N +(-1)^N h_N =\Lambda^N \left( \re^\mm+ \re^{-\mm} \right)+ W_N(\mx)\nonumber\\
\hphantom{\mH_N +(-1)^N h_N}{} = \alpha^{2 \lambda_1} \Lambda^N \widetilde \mm^2 + \sum_{k =0}^{N} (-1)^k \alpha^{(N-k) \lambda_2+ \Delta_k} \widetilde \mx^{N-k} \widetilde h_k + \cdots,
\end{gather*}
where the dots indicate higher order corrections. We now choose $\lambda_1$, $\lambda_2$ and $\Delta_k$, $k=2, \dots, N$, in such a way that
\begin{gather*}
\mH_N +(-1)^N h_N = \alpha^{2\lambda_1} \big( \mH^{\rm QM}_N + (-1)^N \widetilde h_N\big) + \cdots,
\end{gather*}
where
\begin{gather*}
 \mH^{\rm QM}_N= \Lambda^N \widetilde \mm^2 + \sum_{k =0}^{N-1} (-1)^k \widetilde \mx^{N-k} \widetilde h_k.
\end{gather*}
 This requires that
\begin{gather*}
 2 \lambda_1= N \lambda_2, \qquad (N-k) \lambda_2+ \Delta_k = 2 \lambda_1, \qquad k=2, \dots, N,
\end{gather*}
 which, after imposing (\ref{ll12}), is solved by
\begin{gather*}
 \lambda_1= {N \over N+2}, \qquad \lambda_2= {2\over N+2},
\end{gather*}
 and
\begin{gather*}
 \Delta_k = {2 k \over N+2}, \qquad k=2, \dots, N.
\end{gather*}

For $N\ge 3$, the scaling of the $h_k$ is precisely what is found at the AD point of $\CN=2$ Yang--Mills theory with gauge group ${\rm SU}(N)$~\cite{ad,ehiy}. This means that the standard quantum mechanical Hamiltonian with a polynomial potential of degree~$N$ can be obtained by looking at the Omega-deformed $\CN=2$ Yang--Mills theory near the AD point, after taking simultaneously $\hbar \rightarrow 0$. In particular, the standard cubic potential corresponds to the AD point of ${\rm SU}(3)$, while the quartic potential is realized by the AD point of ${\rm SU}(4)$. This is in agreement with the results of~\cite{gg-qm}.\footnote{In the case of monic potentials, the identification between AD points and quantum-mechanical systems was noted in \cite{ito-shu}, which used the ODE/IM correspondence of~\cite{ddt,dt} to study the corresponding spectral problem.} Our quantization conditions, expanded around this point, should give exact quantization conditions for the conventional quantum-mechanical anharmonic oscillators.

However, in order to implement this concretely, we would need analytic expressions for the NS free energy and quantum mirror maps near this point, as a function of the rescaled moduli~$\widetilde h_k$ and~$\hbar_{\rm QM}$. The NS free energy and quantum mirror maps are defined by instanton expansions in gauge theory, which are suited for the region in moduli space where the~$h_k$ are large. The convergence of these expansions near the AD point
(which belongs to the discriminant locus of the SW curve) is by definition very poor. In order to obtain convergent expansions in this and other regions, we would need to perform analytic continuations of the semiclassical expansions. Unfortunately, with the current techniques, these analytic continuations can only be done order by order in the~$\hbar$ expansion.

Another possibility is to find the scaling limit of the TBA system near this point in moduli space. Such a limiting TBA system should lead to the quantization conditions for particular potentials found in~\cite{ddt,dt,
oper} in terms of non-linear integral equations.

\section{Conclusions and outlook}\label{sec-con}

The results presented in this paper open many avenues for further investigation.

Our conjectural quantization conditions have passed many tests, but it would be interesting to prove them, at least at the physical level of rigor. In order to do this, one could develop the analogue of the (exact) WKB method for our deformed theory, and use semiclassical intuition (maybe based on a path integral formulation) to understand the non-perturbative corrections appearing in (\ref{eqc-even}), (\ref{eqc-odd}). More generally, the deformation of quantum mechanics presented in this paper can be regarded as a solvable testing ground for many structures that appear in conventional quantum mechanics. One could study the deformed versions of PT-symmetric Hamiltonians, of quasi-exactly solvable models and of supersymmetric quantum mechanics. One could also apply techniques from the theory of resurgence to understand the non-perturbative structure that we have uncovered. For example, the perturbative series for the energy levels of the Hamiltonian (\ref{dqmh}) can be studied in detail by using a generalization of the BenderWu package \cite{gu-s, bwpack} due to Jie Gu, and the large order behavior
of these series should contain information about the non-perturbative corrections that we have found. Finally, since our exact quantization conditions involve explicit functions, it should be possible to study in detail the analytic continuation of the eigenvalue problem to complex values of the parameters, as in~\cite{bw}.

As we have also emphasized, our deformed version of quantum mechanics displays phenomena which are forbidden in the conventional version. Most notably, we find that, for special values of the parameters, tunneling is suppressed. This leads, in the case of even potentials, to spontaneous parity-symmetry breaking, and in the case of unbounded potentials, to special resonances with real energies. It would be interesting to explore in more detail these qualitatively new phenomena and their implications.

In this paper we have focused on the spectrum of the Hamiltonian, but we would like to determine as well the eigenfunctions. The TS/ST correspondence also provides information on these \cite{mz-wv,mz-wv2} (the results on wavefunctions obtained in \cite{kpamir, sciarappa2} are likely to be valid only at the Toda lattice points, and not in the more generic case we have considered here). Understanding the 4d limit of the proposal of \cite{mz-wv,mz-wv2} is therefore an obvious direction to pursue.

In this paper we have considered the simplest quantum problem arising in SW theory, corresponding to pure ${\rm SU}(N)$, $\CN=2$ Yang--Mills theory. The quantization of SW curves with other gauge groups and/or matter content is certainly possible, and leads to spectral problems which are likely to be solvable by combining the TS/ST correspondence with geometric engineering. This would provide an interesting testing ground for the conjectures put forward in \cite{cgm,ghm}, and would enrich the world of solvable quantum-mechanical models with new acquisitions.

It would be also very interesting to understand in more detail the relation to integrable systems. It was pointed out in \cite{fhm,ggu,huang-blowup,swh} that the spectrum of the integrable systems associated to quantum mirror curves can be obtained by considering the common vanishing locus of va\-rious quantum theta functions, differing among them by a rotation of the moduli. Presumably, the same thing is true in the 4d limit, and by rotating appropriately the Hamiltonians we can find different spectral problems with different quantization conditions, in such a way that the intersecting loci give precisely the Toda lattice points. However, it is not clear to us why this procedure should lead to an enhanced decay of the wavefunctions, as required by the analysis of~\cite{gp}. This enhanced decay also takes place in the case of cluster integrable systems \cite{mz-wv2}. It would be very interesting to clarify this issue in the comparatively simpler case of quantized SW curves.

Perhaps the most pressing problem open by our results is how to make contact with ordinary quantum mechanics. As we have explained, our deformation reduces the problem to understanding the exact NS free energy near the AD points of moduli space. This is not straightforward. However, the fact that TBA formulations of quantization conditions exist for some quantum-mechanical Hamiltonians \cite{ddt,dt,oper} suggests that an analytic determination of this scaling regime is under reach. Such a determination, combined with the limit of our quantization conditions, would provide an exact, resummation-free solution of the anharmonic
oscillator.

\subsection*{Acknowledgements}
We would like to thank Yoan Emery, Giovanni Felder, Matthias Gaberdiel, Jie Gu, Nikita Nekrasov, Massimiliano Ronzani and Szabolcs Zakany for useful discussions.
We are particularly thankful to Jie Gu for extending the BenderWu package to Hamiltonians like the one we study in this paper. The work of M.M.\ is supported in part by the Fonds National Suisse, subsidies 200021-156995 and 200020-141329, and by the NCCR 51NF40-141869 ``The Mathematics of Physics'' (SwissMAP).

\pdfbookmark[1]{References}{ref}
\LastPageEnding

\end{document}